# An Operating Principle of the Cerebral Cortex, and a Cellular Mechanism for Attentional Trial-and-Error Pattern Learning and Useful Classification Extraction

**Marat M. Rvachev***

*Correspondence: rvachev@alum.mit.edu

A feature of the brains of intelligent animals is the ability to learn to respond to an ensemble of active neuronal inputs with a behaviorally appropriate ensemble of active neuronal outputs. Previously, a hypothesis was proposed on how this mechanism is implemented at the cellular level within the neocortical pyramidal neuron: the apical tuft or perisomatic inputs initiate "guess" neuron firings, while the basal dendrites identify input patterns based on excited synaptic clusters, with the cluster excitation strength adjusted based on reward feedback. This simple mechanism allows neurons to learn to classify their inputs in a surprisingly intelligent manner. Here, we revise and extend this hypothesis. We modify synaptic plasticity rules to align with behavioral time scale synaptic plasticity (BTSP) observed in hippocampal area CA1, making the framework more biophysically and behaviorally plausible. The neurons for the guess firings are selected in a voluntary manner via feedback connections to apical tufts in the neocortical layer 1, leading to dendritic $Ca^{2+}$ spikes with burst firing, which are postulated to be neural correlates of attentional, aware processing. Once learned, the neuronal input classification is executed without voluntary or conscious control, enabling hierarchical incremental learning of classifications that is effective in our inherently classifiable world. In addition to voluntary, we propose that pyramidal neuron burst firing can be involuntary, also initiated via apical tuft inputs, drawing attention towards important cues such as novelty and noxious stimuli. We classify the excitations of neocortical pyramidal neurons into four categories based on their excitation pathway: attentional versus automatic and voluntary/acquired versus involuntary. Additionally, we hypothesize that dendrites within pyramidal neuron minicolumn bundles are coupled via depolarization cross-induction, enabling minicolumn functions such as the creation of powerful hierarchical "hyperneurons" and the internal representation of the external world. We suggest building blocks to extend the microcircuit theory to network-level processing, which, interestingly, yields variants resembling the artificial neural networks currently in use. On a more speculative note, we conjecture that principles of intelligence in universes governed by certain types of physical laws might resemble ours.

## 1. Introduction

Despite decades of research and a tremendous amount of gained insight, the operating principles of the cerebral cortex remain unclear (Horton and Adams, 2005; Schuman et al., 2021; Chéreau et al., 2022). The principal neocortical circuit element is the pyramidal cell (PC), comprising at least 70% of the neocortical neuronal population (Nieuwenhuys, 1994). PCs are thought to use three main information integrating spikes: $Na^+$ spikes initiated at the axon hillock, N-methyl-D-aspartate (NMDA) spikes initiated at fine dendritic branches, and dendritic $Ca^{2+}$ spikes (dendritic plateau potentials) initiated near the top of the apical trunk (Figure 1A; Larkum et al., 1999; Williams and Stuart, 1999; Schiller et al., 2000; Larkum et al., 2009; Palmer et al., 2014). In neocortical layer (L) 5 PCs, an excitation coupling between the $Ca^{2+}$ and $Na^+$ spike initiation zones has been observed. In this coupling, the dendritic $Ca^{2+}$ spike initiation threshold is dramatically lowered by a single $Na^+$ backpropagating action potential (bAP), such that a small $Ca^{2+}$ initiation zone depolarization evolves into a broad (>10 ms) dendritic plateau potential, concurrently with burst firing of multiple high-frequency $Na^+$ spikes. Such $Ca^{2+}$ spike firing is termed backpropagation activated $Ca^{2+}$ spike firing (BAC firing; Figure 1B), and it creates coincidence detection between the tuft and soma compartments (Larkum et al., 1999; Larkum and Zhu, 2002). In addition, the spatial input integration at fine dendritic branches leads to a multidimensional sigmoidal thresholding in the PC input-output function (Figure 1C; Jadi et al., 2014). Several intricate PC plasticity mechanisms have been uncovered, such as spike timing-dependent plasticity (STDP; Markram et al., 1997; Abbott and Nelson, 2000) and, in the hippocampus, behavioral time scale synaptic plasticity (BTSP; Bittner et al., 2015; Bittner et al., 2017; Grienberger et al., 2017). Still, the computational principles that allow learning in neurons to result in cortical function are poorly understood.

In an attempt to address these questions, a cellular mechanism for cortical associations has been suggested and has attracted considerable interest (Larkum, 2013). In this model, the coincidence detection between the PC tuft and soma compartments, as well as BAC firing, endows the cortex with an inbuilt associative mechanism for combining feedback (internal, prediction) and feedforward (external, sensory) information arriving at the two compartments. The main function of the cortex is presumed to be the association of external data with an internal representation of the world, which is achieved at the cellular level: feedback serves as the cortex's "prediction" of whether a particular PC could or should be firing, but only if that neuron receives enough feedforward input does BAC firing take place. Alternative models for cortical function have also been proposed (Mumford, 1992; Rao and Ballard, 1999; Friston, 2010; Bastos et al., 2012; Gilbert and Li, 2013; Grossberg, 2013; Hawkins and Ahmad, 2016; Heeger, 2017; Keller and Mrsic-Flogel, 2018; Hawkins et al., 2019; Richards and Lillicrap, 2019; Bennett, 2020; Zagha, 2020).

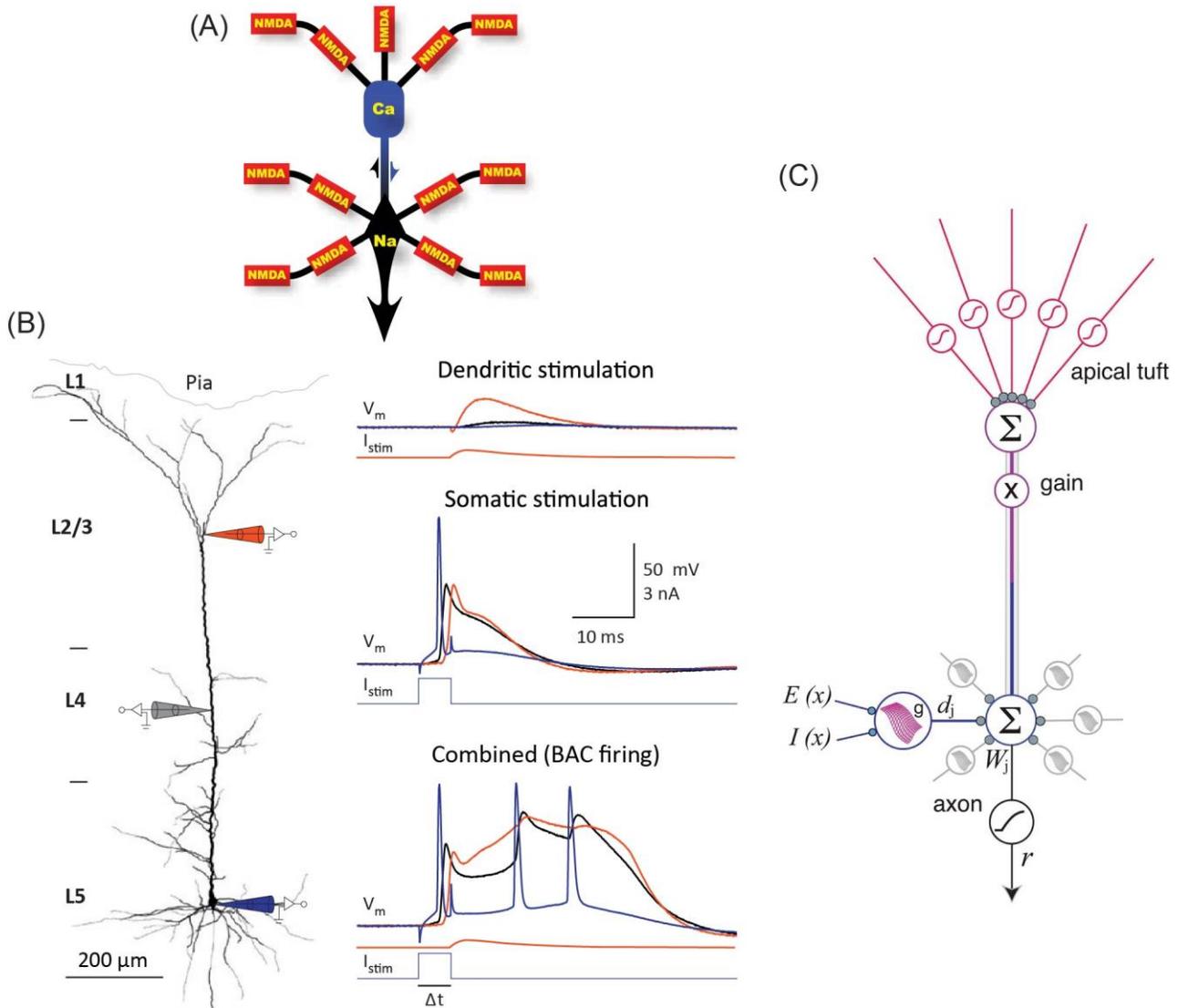

**Figure 1. Information integration in PCs. (A)** Schematic diagram of the main information integrating subcompartments in a typical layer (L) 5 PC, with the corresponding spikes noted. NMDA spikes nonlinearly sum synaptic inputs in semi-independent subcompartments in the fine apical tuft and basal dendrites. The resulting output is nonlinearly integrated in the Na$^+$ and Ca$^{2+}$ initiation zones, which influence each other (as indicated by arrows). Adapted, with permission, from Larkum et al. (2009). **(B)** Left panel. Schematic representation of an experiment on an L5 PC in vitro, with electrode positions depicted symbolically. Electrode color scheme is the same as for the current injected, I$_{stim}$, and potential recorded, V$_m$, shown in the right panel. Right panel. Top: Dendritic current injection simulating an excitatory postsynaptic potential (0.3 nA amplitude) resulted in no spikes and minimal impact at the soma. Middle: Threshold current injected at the soma evoked a single Na$^+$ spike and a smaller amplitude but longer duration response in the dendrite. Bottom: BAC firing; the same two current injections separated by 5 ms evoked a broad dendritic Ca$^{2+}$ spike accompanied by a burst of Na$^+$ spikes, as the initial backpropagating Na$^+$ spike lowered the Ca$^{2+}$ spike initiation threshold. Adapted, with permission, from Larkum et al. (1999) and Larkum (2013). **(C)** A model of information integration in PCs. Dendritic subunit outputs d$_j$ are summed with weights W$_j$ and thresholded at the axon hillock, and themselves have multidimensional sigmoidal input-output functions g that depend on the spatial patterns of excitation E(x) and inhibition I(x). Apical tuft output sets overall gain for the cell. Adapted, with permission, from Jadi et al. (2014).

This paper proposes that one of the functions of the excitation coupling between the PC tuft and soma compartments is to enhance learning efficiency in the "pyramidal neuron as a reward-modulated combinatorial switch" scheme (Rvachev, 2013). This framework is grounded in the concept of learning by trial and error with reinforcement of favorable outcomes and weakening of unfavorable ones, known as the "law of effect" (Thorndike, 1898), and in the model of PC as a two-layer neural network (Poirazi et al., 2003). In this framework, either apical tuft or perisomatic PC inputs initiate "guess" neuron firings, while the excitation strength of basal synaptic clusters, which encode incident input patterns, is adjusted based on reward feedback. Although the framework is straightforward, it enables PCs to function in a remarkably intelligent manner, primarily because PCs serve as classifiers in our inherently classifiable world.

We modify the plasticity rules in Rvachev (2013) to be closer to the experimentally observed BTSP, which preserves the results of that paper while making the framework more biophysically and behaviorally plausible. We postulate that neocortical PC dendritic plateau potentials and associated burst firing are neural correlates of attentional, aware processing, at least in certain types of PCs, generally consistent with experiments on consciousness and perceptual detection (Takahashi et al., 2016; Suzuki and Larkum, 2020; Takahashi et al., 2020), and we categorize various PC excitations as attentional versus automatic and voluntary/acquired versus involuntary, based on their excitation pathways. We suggest voluntary attentional excitations are driven by feedback corticocortical (CC) and cortico-thalamo-cortical (CTC) L1 projections, while involuntary attentional excitations are driven by medial temporal lobe (MTL) and subcortical L1 projections, to attend to important cues such as novelty, discrepancy, and noxious stimuli. Voluntary attentional excitations become automatic with learning, while involuntary attentional excitations do not, due to the wiring architecture. Further, we propose mechanisms for cross-layer PC cooperation through depolarization cross-induction in the minicolumn dendritic bundles and suggest building blocks for a broader network-level application of the microcircuit theory. On a more speculative note, we hypothesize that the nature of intelligence in universes governed by certain types of physical laws might be similar to ours.

## 2. The Extended "Pyramidal Neuron as a Combinatorial Switch" Model

The learning problem posed for the organism, or a part of its nervous system, can be stated as follows: given an arbitrary pattern X of active neuronal inputs, find and memorize a corresponding "optimal" pattern Y* of active neuronal outputs (Figure 2A). The optimal pattern is defined as one that produces an action resulting in a positive reward, or at the very least, the absence of a negative reward for the organism; generally speaking, the pattern can be arbitrary from a combinatorial perspective. Rewards are mediated by signaling molecules, such as neuromodulators or hormones. Note that in Rvachev (2013) and in this paper, a "negative" reward refers to a punishment, while a "positive" reward is a reward per se. The only minor difference in this formulation from Rvachev (2013) is the postulation that the absence of rewards, which is expected to be a relatively infrequent occurrence, is still considered an optimal outcome, to align the model mechanisms with BTSP (discussed later).

The output neurons are PCs that can be arranged in a layer, with their basal dendrites extending in a plane into which the inputs being associated diffusely and potentially randomly project, such that the inputs impinge on the PCs in a similar, although not necessarily identical, manner (Figure 2B). This configuration should enhance the unbiased, arbitrary X to Y learning efficiency, as each PC would have approximately similar capability to learn how to respond to an arbitrary X. For effective learning, the PCs should drive complementary or alternative actions, such as coarse, fine, or alternative movements of a hand in the motor cortex, or choosing between alternative paths when solving a math problem in higher-order cortical areas. For the sake of simplicity, and unless stated otherwise, the learning PCs are assumed to be the neocortical pyramidal tract (PT) PCs, which are thick-tufted, predominantly target subcortical areas, and are situated mostly in the lower L5 (L5b). However, we assume that analogous mechanisms may operate in the intratelencephalic (IT) PCs, which are mostly thin-tufted, project to the cortex and striatum, and are located in layers 2-6 (Hattox and Nelson, 2007; Harris and Shepherd, 2015; Kim et al., 2015; Shepherd and Yamawaki, 2021). We do not consider the corticothalamic (CT) PC type.

Learning proceeds in a trial-and-error fashion. Feedforward and some feedback (context) inputs project on the PT PC basal dendrites (Figure 2B). PC combinations are selected for firing via L1 tuft feedback input (discussed later). The PCs may be brought to fire with the assistance of an "action initiation" mechanism, such as a uniform enhancement in the PC tuft-to-soma coupling via higher-order thalamic input to the cortex (Suzuki and Larkum, 2020), or a depolarization via excitation or removal of inhibition uniformly to the PCs' perisomatic or apical regions, e.g., until a certain level of aggregate PC output activity is achieved. As a result of the tuft activation, the selected PCs burst fire in conjunction with dendritic plateau potentials, which can be achieved with or without BAC firing (Larkum et al., 1999; Francioni and Harnett, 2022). The PCs' basal synapses form clusters that are nonlinearly excited via the NMDA spike mechanism if they receive coincident input. If a cluster excitation is accompanied, within a behavioral time scale (up to seconds), by the PC burst firing, the learning rule is used: long-term enhance (weaken) the cluster excitation strength if the accompanying reward is positive or insignificant (negative), as depicted in Figure 2C. Greater absolute rewards should cause greater learning, as suggested in the "law of effect": "The greater the satisfaction or discomfort, the greater the strengthening or weakening of the bond" (Thorndike, 1898). After a number of learning trials, the clusters that accumulate enhanced strength will be those related to the input subpatterns that predict a positive or insignificant reward and the absence of a negative reward when the neuron fires (Figure 2D). Then, for an arbitrary input pattern, the PCs with a greater number of these enhanced clusters active should have more basal and, hence, somato-axonal excitation than others. The readout of this memory can proceed without the participation of tuft input if the increased somato-axonal excitation is sufficient to fire the PCs, or, if it is not, a mechanism similar to the "action initiation" described above can be employed (Figure 2E).

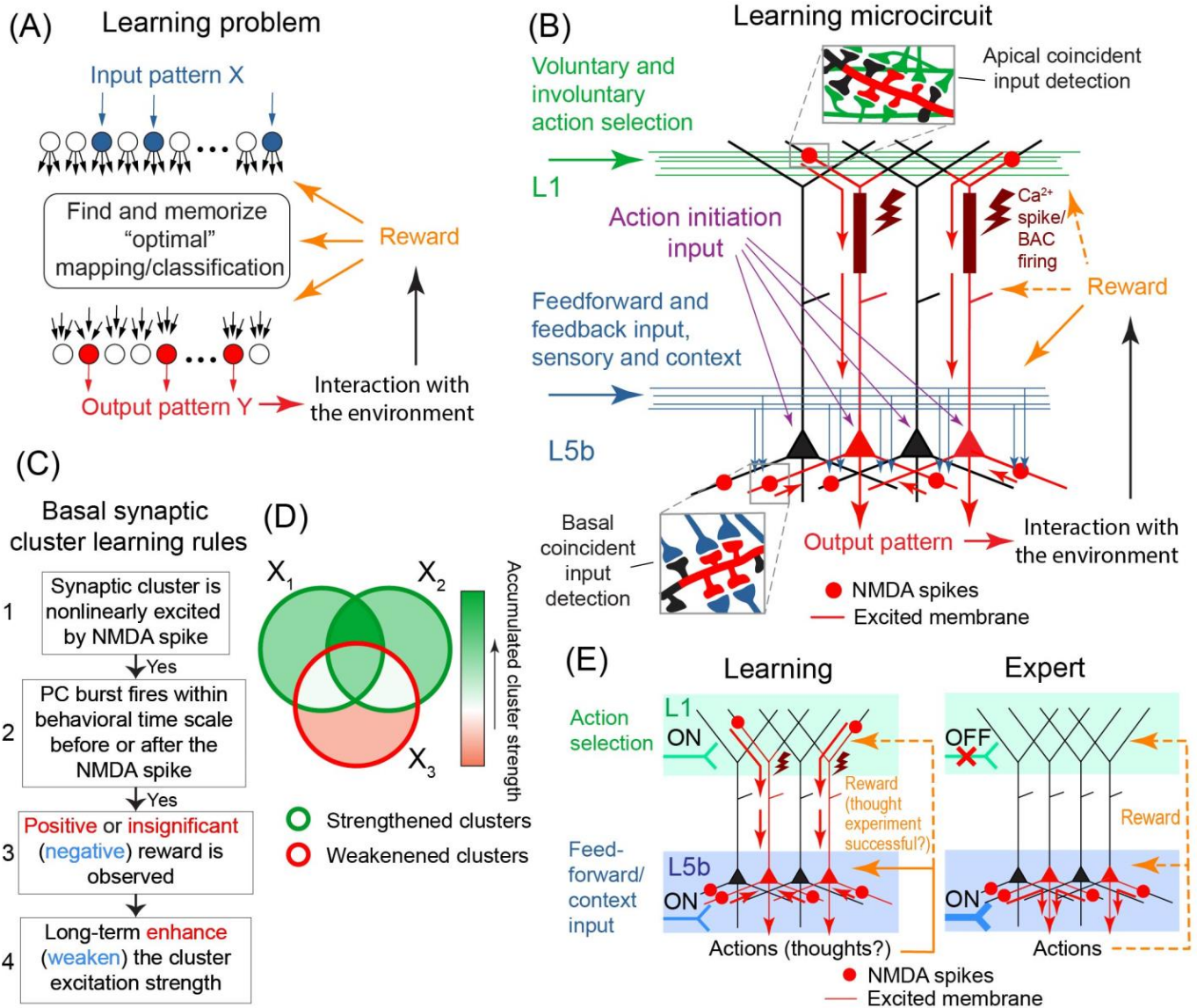

**Figure 2. The learning problem and the proposed microcircuit of the solution. (A)** Schematic representation of the learning problem setup for a neural network. For an arbitrary input pattern X, which may include internal organism state inputs, the objective is to find and memorize an "optimal" response Y* such that its interaction with the environment results in a positive reward or, at the very least, avoids a negative one. We will argue later that the optimal response likely requires a form of input classification. **(B)** Schematic depiction of the proposed microcircuit of the solution. Neocortical pyramidal tract (PT) PCs. Some PCs are selected for firing through L1 inputs, either voluntarily (via CC and CTC projections, discussed later) or involuntarily (reflecting attention-drawing cues such as novelty, discrepancy, noxious stimuli, etc., also discussed later), which results in dendritic Ca²⁺ spikes (brown rectangles and flashes) with PC burst firing (red PCs). Feedforward sensory and some feedback (context) inputs form basal synaptic clusters. Spatiotemporal input coincidence in these clusters results in NMDA spikes (red circles), which reflect the situation the organism is in. The long-term excitation strengths of the clusters are modified based on the learning rules in **(C)**. Following learning, PCs can execute actions based on the combinatorial memory **(D)** and **(E)**. For the "action initiation" mechanism, only the perisomatic pathway is shown (purple arrows; see text for details). Red arrows indicate the general direction of signal propagation in PCs. **(C)** Diagram of the proposed learning rules for basal synaptic clusters. The rules are suggested to be similar to BTSP, with an added modulation by rewards and potentially different time course of underlying processes such as eligibility traces. Generally, steps 1 to 3 are presumed to be present for learning to occur. However, in some well-functioning systems, steps 2 and/or 3 might not be present. If step 3 is absent, the cluster strength is likely to be enhanced in step 4. Overall PC excitability may be regulated by homeostatic renormalization of synaptic strength. **(D)** Schematic representation of the accumulation of synaptic cluster strength during learning. The areas X₁, X₂, and X₃ represent sets of clusters modified by three distinct training episodes. One-shot learning is also possible. **(E)** Left panel. A more concise rendering of **(B)**. As depicted, the output pattern may represent actions as well as thoughts, such as guesses about a classification. Right panel. An expert (well-trained) organism executes memorized actions, as described in the text. The area X₁, X₂, and X₃ represent sets of clusters modified within that layer. The input input into L5b symbolizes strengthened synaptic clusters within that layer. Throughout the figure, orange dashed arrows depict reward delivery pathways that are allowed, but not required within the model. Design of **(E)** is inspired by Doron et al. (2020). Figure 2 represents a modification and extension of the model in Rvachev (2013).

The suggested synaptic plasticity rules (Figure 2C) are quite similar to BTSP, although the role of neuromodulators or hormones (which could represent rewards) within BTSP is not yet clear. BTSP has been observed in the formation of place fields in the hippocampal area CA1 (O'Keefe, 1976; Bittner et al., 2017; Magee and Grienberger, 2020). CA1 PC basal trees receive a constant barrage of excitatory inputs (Bittner et al., 2015; Grienberger et al., 2017; Davoudi and Foster, 2019), while a local population of inhibitory interneurons produces inhibition balancing out this excitation (Grienberger et al., 2017), thus resulting in no place-selective firing in "silent cells." If a plateau potential is initiated in the distal tuft of such a cell, it spreads throughout the cell and induces a potentiation of excitatory basal inputs that arrive around the plateau (Bittner et al., 2015; Bittner et al., 2017), long-term enhancing the inputs weights and producing place field firing (Harvey et al., 2009; Epsztein et al., 2011; Bittner et al., 2015; Diamantaki et al., 2018). The enhancement of weights is thought to be gated by eligibility traces generated by NMDA spikes, evoked by either neighboring (i.e., clustered) synapses or repetitive stimulation of individual synapses (Magee and Grienberger, 2020). According to our model an analogous mechanism, likely inclusive of the inhibitory interneuron effects, operates in the neocortex, but it may also be modulated by rewards as postulated here, and may exhibit different time symmetry properties compared to the time-asymmetric BTSP. We consider the putative NMDA spikes resulting from repeated stimulation of or a potent single synapse, as a specific case within the broader category of heterogeneous clusters. The induction of BTSP, both in vivo and in vitro without the delivery of rewards, and in vivo with positive rewards (Bittner et al., 2017), aligns with the present model formulation. It should be noted that neuromodulators are released in all layers throughout the cortex via long-range projections (Schuman et al., 2021), supporting the present framework. The STDP-like plasticity rules in Rvachev (2013) can be substituted with the BTSP-like rules described here without affecting results in that paper.

Although the learning framework described above might appear unsophisticated, it was demonstrated in silico (Rvachev, 2013; Rvachev, 2017) that such a system, with only 12 input and 3 output neurons, exhibited key characteristics of intelligence: it extracted essential differences between object classes ("apples" and "stones," by differentiating features "smooth surface" versus "rough surface" and "stem on top" versus "no stem on top"), despite the presence of identical features (symmetrical and rounded shapes), while disregarding varying features (sizes and colors). Effectively, the system formulated abstract concepts of the classes, enabling it to correctly classify unencountered specimens (eating apples, discarding stones and never "doing nothing") for over 97% of random learning object sequences, and it would likely appear intelligent to an external observer. Figure 3 illustrates the process of how abstract concepts are distilled within cluster weights in the present framework, using a simplified example with 2-synapse clusters. On the other hand, the framework also performs well in larger-scale memorization tasks in a sparse coding regime (Rvachev, 2013).

The fundamental reason the present learning scheme can be successful in our world is that PCs within the scheme act as efficient classifier units, operating on combinations of input features. On the other hand, our world is indeed highly classifiable, a fact that ultimately stems from the vast number of elementary particles and the invariance of their interactions in space and time (Figure 4A). "A wolf that has learned how to catch a rabbit is more likely to catch another rabbit, as well as another alike animal, in a similar terrestrial environment" (Rvachev, 2013). The scheme should be very successful in environments where, among varying organism inputs, there are specific subpatterns for which there exist actions that generate positive rewards. The organism attempts to find and memorize these subpatterns in the neurons responsible for producing the actions. Conversely, the scheme would not function well in a highly non-classifiable world (Figure 4B).

Theoretically, if other universes exist with a large number of elementary particles interacting under laws different from those of the Standard Model, but still invariant in space and time (or other analogous dimensions), these universes could also be classifiable (Figure 4C). Therefore, if life and intelligence exist in these other universes, we propose that the nature of intelligence could adhere to the suggested principles we observe in our own, i.e., the classification of the external world according to its relevance for the organism.

# 3. Combinatorics of PC Selection via L1 Input

This paper suggests that both voluntary and potentially some types of involuntary selection of PCs for firing are driven by tuft NMDA spikes, which are summed at the $Ca^{2+}$ spike initiation zone at the top of the apical trunk (Figure 2B). A dendritic segment capable of generating NMDA spikes may contain several hundred synapses, and activation of 8-20 of them, depending on their proximity, may be necessary for spike generation (Major et al., 2008; Major et al., 2013). As an example, consider I = 25 input neurons forming clusters of C = 8 synapses, arranged in different permutations on various PC tufts. If G = 8 active synapses are required to generate an NMDA spike in a cluster, this allows for the individual activation of $\binom{25}{8} \approx 10^6$ different PCs, assuming a single NMDA spike can activate a PC. For I = 50, C = G = 20, this number increases to $5 \times 10^{13}$. Therefore, in principle, a very large number of PCs can be individually selectable using combinations of a relatively small number of inputs. It seems plausible that the selection process uses a grid-like architecture (Figure 5). We posit that PCs can be voluntarily selected for firing from both neighboring and distal deep cortical areas, such as by prefrontal cortices controlling arbitrary motor movements via long-range projections.

# 4. Generalized Action/Attention as PT PC Dendritic $Ca^{2+}$ Spikes and PT PC Excitation Classification

In the present model, a PC dendritic $Ca^{2+}$ spike (with associated burst firing) is necessary and sufficient for its learning, given adequate basal inputs and certain rewards. Interestingly, attention (in the sense of awareness) is also evidently necessary and sufficient for the type of higher animal learning discussed (semantic, classification), provided it is accompanied by novelty and thus potential reward (Figure 6A). Hence, under the present framework, it is reasonable to suggest that attention (at least in the neocortex) is directly tied to both IT and PT PC dendritic Ca²⁺ spikes across various layers. However, recent studies have found a link between dendritic Ca²⁺ spikes specifically in L5 PT (not IT) PCs and suprathreshold perceptual detection (Takahashi et al., 2016;

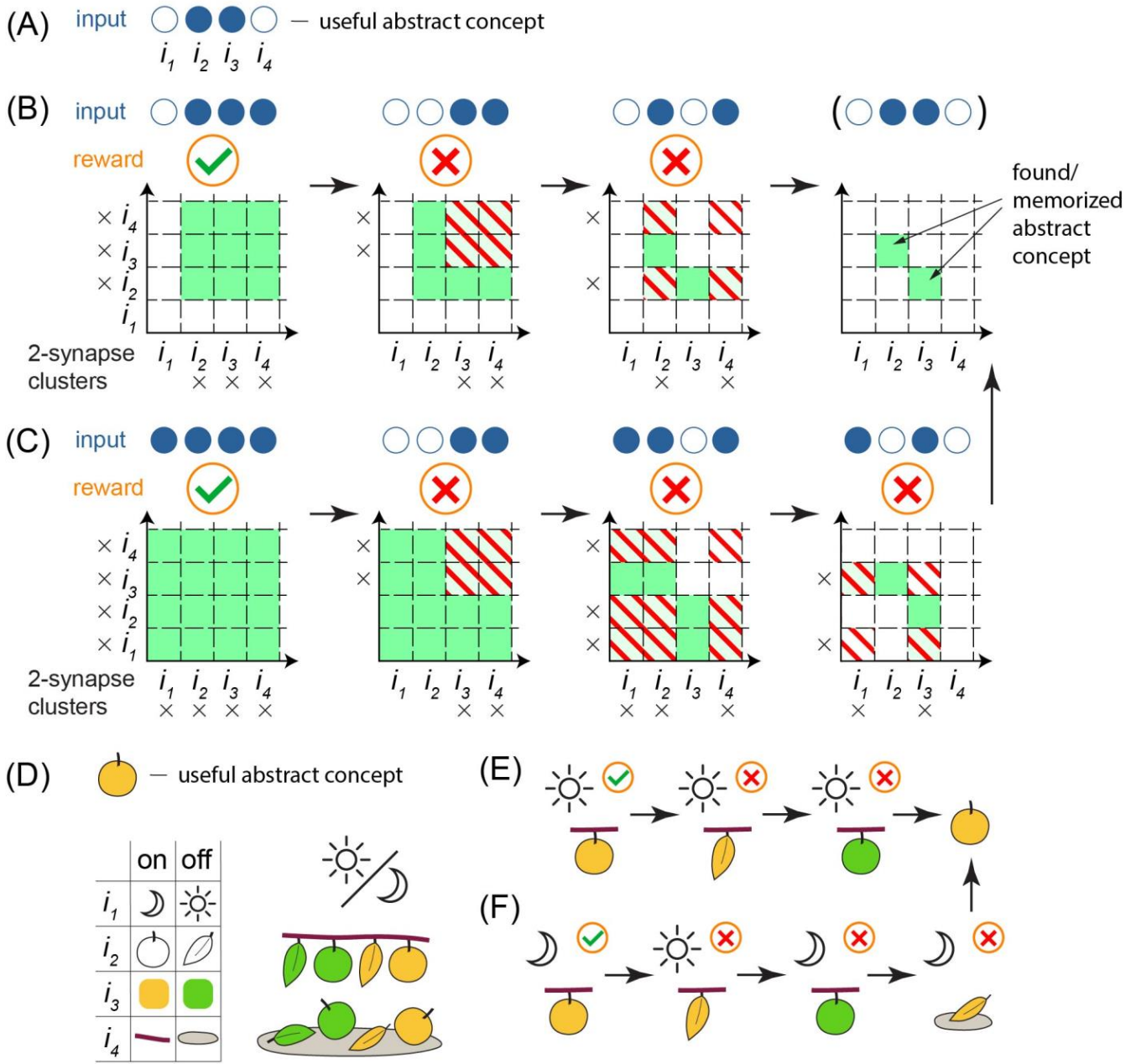

**Figure 3. A simplified illustration of a PC extracting abstract concepts within the present framework. (A)** As an example, a useful abstract concept for a PC that classifies objects as edible is defined as the active combination of inputs 2 and 3 out of a total of 4. Objects exhibiting this concept are edible, others are not. This concept is abstract, as it is assumed that no object exists with only these inputs active. Filled blue circles represent active inputs, empty blue circles indicate inactive ones. **(B)** A learning sequence leading to the deduction of the concept. Orange circles below the inputs indicate the reward (a green checkmark for a positive reward when the object contains the concept, a red cross for a negative reward otherwise) following the PC's trial firing. Squares in the grid represent individual basal clusters, each consisting of 2 synapses, one for each input on the axes. Black crosses mark active inputs. White squares represent clusters with unenhanced strength, green denotes enhanced strength clusters, stripes indicate the removal of strength enhancement (if present) due to the negative reward associated with the cluster excitation and PC trial firing. In this sequence, input 1 is always inactive, while input 4 is always active. The presentation of an object possessing the useful abstract concept, followed by two objects that lack the full concept but contain some of its active inputs, results in the correct deduction of the concept. Note that it is common practice, when learning a rule with exceptions, to state the general rule first, followed by its exceptions, analogous to the learning sequence here. **(C)** Another learning sequence leading to the deduction of the concept. Here, inputs 1 and 4 are not always active or inactive, necessitating an additional learning step, as the variations in inputs 1 and 4 create possibilities for other valid concepts. **(D)-(F)** A reinterpretation of **(A)-(C)** with specific features assigned to the inputs. **(D)** Top panel. The useful abstract concept is a ripe (yellowish) apple, for a PC classifying objects as edible. Bottom panel. Left: the depicted definition of input states is based on the time of day (night, day), object shape (round, elongated), color (yellowish, green), and location (branch, ground). Right: possible situations for encountering these objects. **(E), (F)** Training sequences equivalent to **(B)** and **(C)**, with trial eating of the object, resulting in the correct deduction of the abstract concept as a ripe apple.

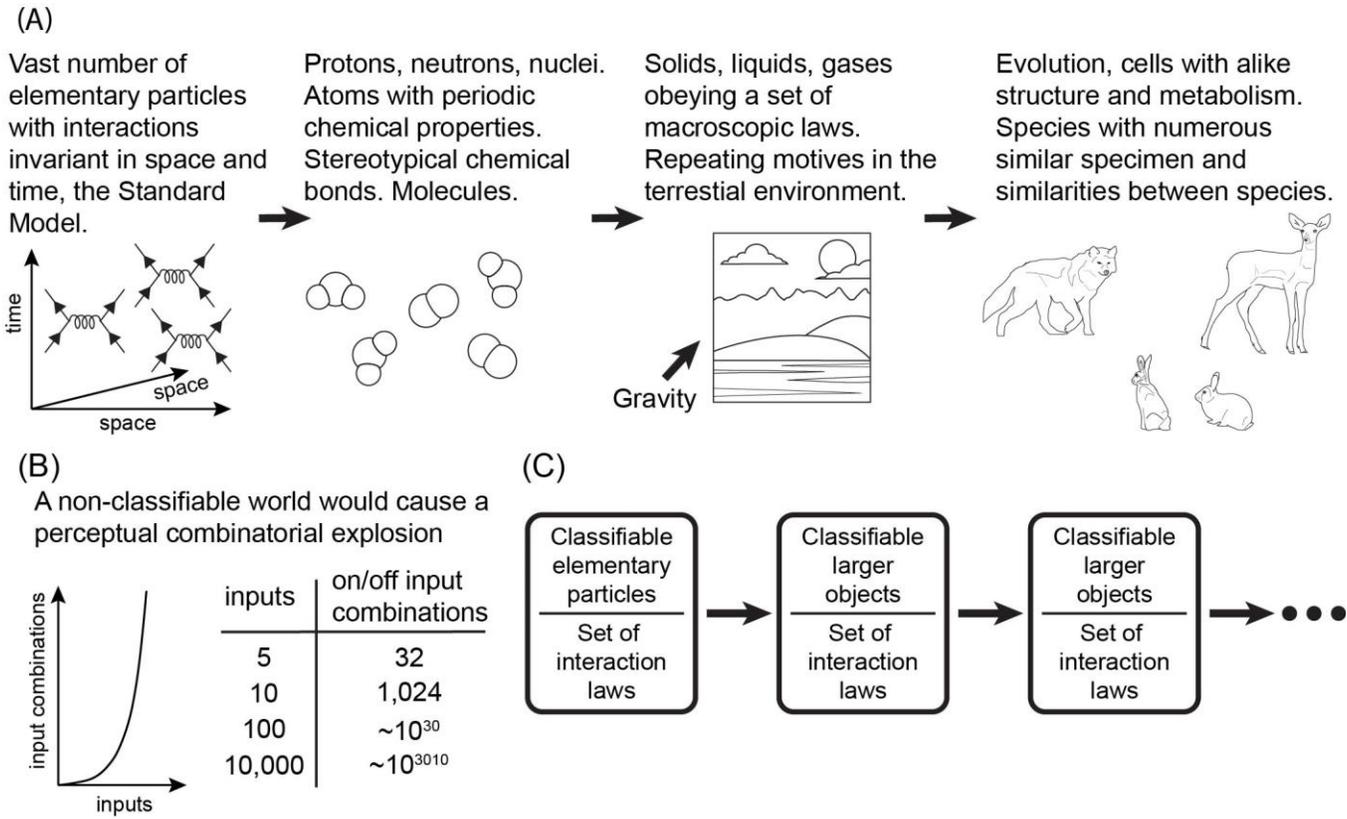

**(A)**

Vast number of elementary particles with interactions invariant in space and time, the Standard Model.

Protons, neutrons, nuclei. Atoms with periodic chemical properties. Stereotypical chemical bonds. Molecules.

Solids, liquids, gases obeying a set of macroscopic laws. Repeating motives in the terrestial environment.

Evolution, cells with alike structure and metabolism. Species with numerous similar specimen and similarities between species.

Gravity

**(B)**

A non-classifiable world would cause a perceptual combinatorial explosion

| inputs | on/off input combinations |
|---|---|
| 5 | 32 |
| 10 | 1,024 |
| 100 | $\sim 10^{30}$ |
| 10,000 | $\sim 10^{3010}$ |

**(C)**

Classifiable elementary particles / Set of interaction laws → Classifiable larger objects / Set of interaction laws → Classifiable larger objects / Set of interaction laws → • • •

**Figure 4. (A)** Schematic diagram illustrating how the high classifiability of the world at the macroscopic scales ultimately stems from the vast number of a limited set of types of elementary particles/fields and the invariance of the laws of physics in space and time. **(B)** Schematic graph and table illustrating the scale of the combinatorial explosion that learning systems would have to contend with in a hypothetical non-classifiable world. **(C)** The classifiability of our world at macroscopic scales seems to be independent of the particular form of interactions in the Standard Model. In fact, it appears to be a general property of universes that abide by laws invariant in space and time, or their analogous dimensions.

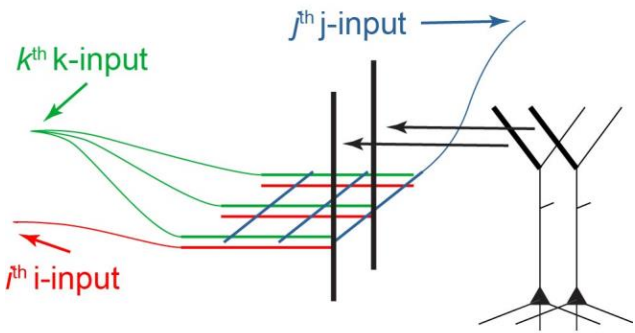

**Figure 5. Schematic diagram of a hypothetical grid-like architecture for the selection of PCs for firing via tufts.** PCs are arranged in a 2D sheet, as in a cortical layer (only 2 PCs are depicted). The I "i-inputs," J "j-inputs," and K "k-inputs" target PT tufts, creating clusters with unique i-j-k combinations, which allows for the arbitrary selection of one out of IJK PCs, assuming that 3 inputs are sufficient to excite a cluster and a cluster can excite a PC. Different k-inputs target different cortical regions (only one cortical region is shown). This process may be able to select multiple PCs sequentially if an excitation maintenance mechanism can preserve the excitation of selected PCs. This architecture can be extended to more input dimensions, generally permitting the selection of a larger number of PCs using the same number of inputs.

Takahashi et al., 2020), which we interpret as a form of attention. Consequently, here we adopt a view that PT (and not IT) PC dendritic Ca²⁺ spike is a neural correlate and manifestation of neocortical attention. Nevertheless, there remains the possibility that an IT PC-associated attention phenomenon was present but not detected in the aforementioned studies, for instance, if it was linked to an internal memory of a reward presentation rather than the perception of the reward presence per se. In any case, within this framework, it seems plausible that the subjective experiences of attention, awareness, and consciousness are also associated with subcortical structures that evolved before the neocortex and may be recruited by neocortical PC dendritic Ca²⁺ spikes with burst firing, as has been previously discussed (Llinás et al., 1998; Parvizi and Damasio, 2001; Krauzlis et al., 2013; Takahashi et al., 2020).

It should be noted that such defined attention can, in principle, be directed (for example, voluntarily, using "free will") towards any L1-activatable PT PC within the fairly uniform neocortical gray matter. In the motor cortices, this corresponds to the execution of motor actions, which is typically distinguished from attention. Therefore, it is more appropriate to refer to PT PC plateau potentials and associated burst firing as "attentional action and attention," or simply "generalized action/attention."

### 4.1. Voluntary Attentional PT PC Excitation

We suggest that all voluntary, nonroutine, attentional actions, such as (a) deliberate motor actions, (b) visualizing or actively thinking about a situation, problem, or its solution, (c) imagining an object, and (d) focusing attention on a sensory sensation or a pattern of sensations, are executed as L1-activated PT PC firing. For example, if a human is instructed to perform a nonroutine action, such as pointing their left pinky finger downward, imagining an orange elephant, or focusing attention on a particular sound in an auditory stream, we posit that the corresponding PT PCs will be selected for activation via L1 input. These actions belong to the domain of the generalized action/attention introduced above.

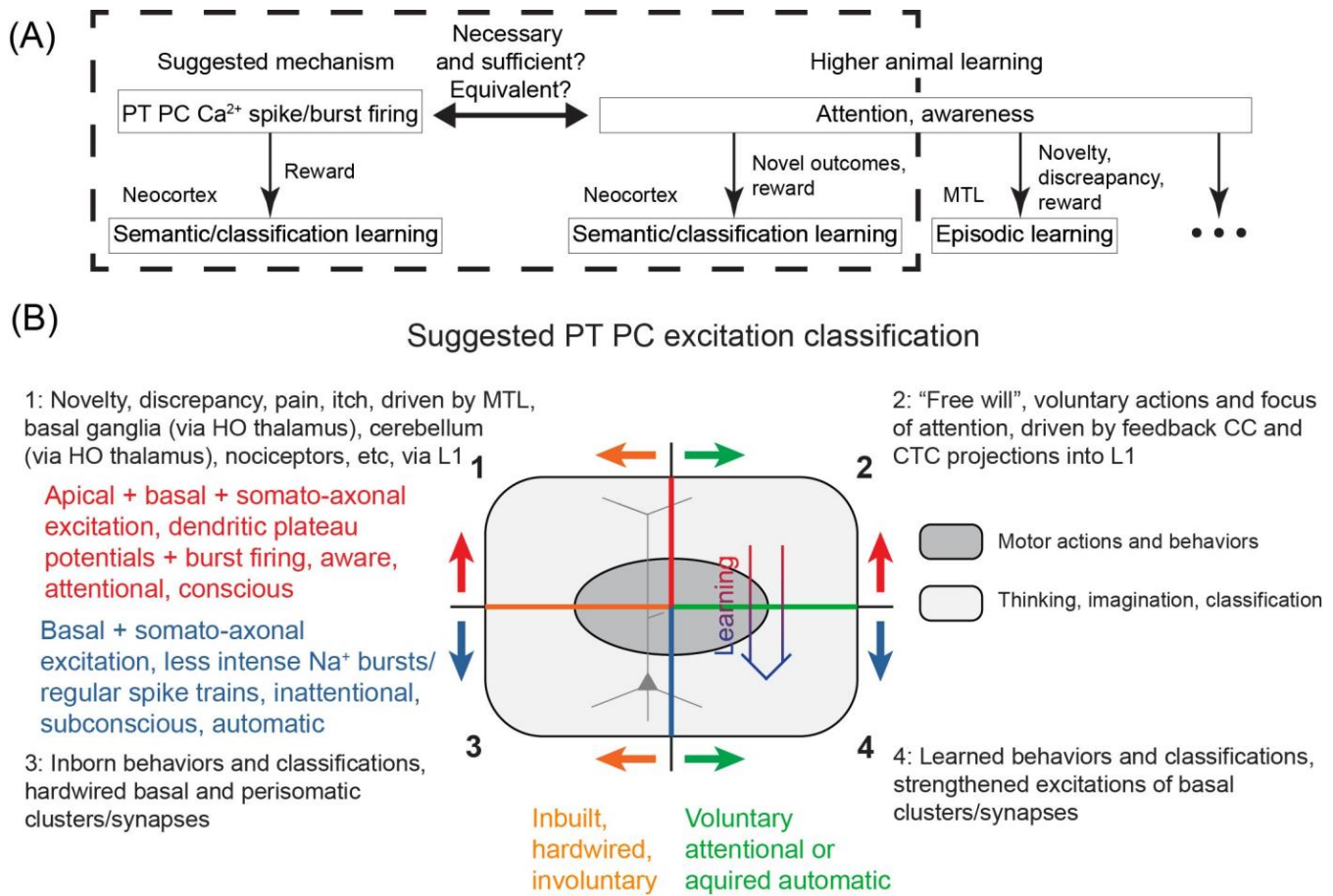

**Figure 6. (A)** Schematic diagram illustrating the similarity, and possible equivalence, between PT PC dendritic Ca²⁺ spike with burst firing in the discussed framework and attention/awareness in higher animal learning, in that they are necessary, but still require some form of rewards for learning. The diagram attempts to separate neocortical learning from other types of learning that might also encompass what is subjectively perceived as attention and awareness. While attention and awareness have slightly different connotations, this distinction is neglected in the present study. **(B)** Suggested PT PC excitation classification, described in the text. Voluntary attentional motor actions and thoughts/classifications (quadrant 2) can transition into acquired automatic excitations (quadrant 4) with learning (thick arrow), while involuntary attentional excitations (quadrant 1) are hardwired into L1 and cannot become automatic. Superimposed is a schematic depiction of a PC, with its apical dendrite in the upper half, corresponding to the apical excitations occurring in the upper quadrants. HO, higher-order.

## 4.2. PT PC Excitation Classification

Given several possible types of PT PC excitations, a classification scheme is proposed in Figure 6B. The top two quadrants describe the powerful dendritic plateau potentials accompanied by burst firing, which are initiated via L1 inputs and correspond to attentional, aware, conscious mental processes. These excitations can either be voluntary, as described above and also suggested to be driven by feedback CC and CTC L1 projections (top right quadrant), or they can be involuntary, resulting from hardwired mechanisms drawing attention to cues such as novelty, discrepancy, pain, itch, or unexpected touch, and driven by MTL and subcortical projections into L1 (top left quadrant, discussed later).

The bottom two quadrants describe less intense burst and regular Na⁺ spiking, which do not require L1 input and correspond to inattentional, subconscious, automatic actions. These can either be inborn, representing inbuilt behaviors and classifications (bottom left quadrant, e.g., recognition of lines in the primary visual cortex (Hubel and Wiesel, 1962)), or learned (bottom right). The oval in Figure 6B schematically represents motor actions and behaviors, which are distinguished from other actions such as thinking, imagination, and classification.

## 4.3. Involuntary Attentional PT PC Excitation

As mentioned above, this paper hypothesizes that the feedback CC and CTC L1 projections (Felleman and Van Essen, 1991; Schuman et al., 2021; Shepherd and Yamawaki, 2021) are the means through which higher-order cortices drive voluntary actions and attention in lower-order cortices. In many currently accepted models, the CC and CTC feedback projection motifs are thought to inform lower-order cortices about the outcome of their activity, or about expected or predicted inputs (Larkum, 2013; Keller and Mrsic-Flogel, 2018; Roelfsema and Holtmaat, 2018). In some models, dopamine and other neuromodulators encode reward prediction error (RPE), which measures whether the outcome of an action is better or worse than expected, with the feedback and RPE-related signals together modulating synaptic plasticity (Schultz, 2017; Roelfsema and Holtmaat, 2018); see (Chéreau et al., 2022) for a review. We suggest that at least a part of these "prediction" and "prediction error" functions, where they occur, are performed via the pathways between the neocortex and the MTL (includes the hippocampus, amygdala, and parahippocampal regions).

The MTL and the hippocampus, which is evolutionarily and anatomically related to the olfactory cortex, could, in one of their roles, serve the function of monitoring the evolution of incoming signals from the neocortex, essentially "smelling" the situation, and comparing it to previous similar episodic memories. If a significant difference (i.e., novelty, discrepancy, error, a new object in a scene) is detected, the MTL could direct attention to it by burst firing the corresponding neocortical PT PCs via their L1 inputs using the reciprocal looping circuits with the neocortex (Figure 7A), the same circuits that are thought to implement neocortical recall of episodic memory (Rolls, 2010; Figures 7B,C). This attention could be raised in a local neocortical area, not necessarily a particular PT PC or cortical minicolumn, by sequential trial firing of local PCs, aimed at classifying the novel aspect of the situation, e.g., into behavioral categories such as "feeding opportunity," "threat," etc., using appropriate exploratory behaviors.

This scheme is consistent with the observation that suppression of the perirhinal cortex (PRh, the last station in the MTL loop projecting to a part of the neocortex, Figures 7B,C) input in L1 of the rodent primary somatosensory cortex (S1) disrupted learning to associate direct electrical microstimulation of S1 L5 PCs with a reward (Doron et al., 2020). In the present framework, this is interpreted as the inability of the MTL to draw attention to the novel stimulus via L1 input after the MTL detects it (Figure 7, blocked red dashed pathways). Essentially, for sensory stimulation to attract attention, the MTL must recognize it as a novel stimulus and therefore worth paying attention to. On the other hand, it was reported that expert (well-trained) animals were able to behaviorally react to the stimulus in the learned fashion after the PRh suppression. This is interpreted as the basal and somato-axonal execution of the learned actions, which does not require L1 input (Figure 6B, quadrant 4). It was also reported that bursts, but not regular spike trains, retrieved learned behavior. Within the present framework, automatic/inattentional learned behaviors can be driven by Na$^+$ burst firing without plateau potentials, similar to the recall-associated bursts in the hippocampal place cells (Bittner et al., 2015; Bittner et al., 2017; Milstein et al., 2021). It is also plausible that automatic behaviors are more difficult to achieve in the primary sensory areas and additional voluntary focus of attention is needed to generate behavior. Although apical dendritic excitability was enhanced after learning (Doron et al., 2020), we suggest that this is not the primary learning mechanism.

Moreover, we suggest that BTSP-induced memory in CA1 PCs can also raise attention in the neocortex, via the neocortical-MTL looping circuits, in near-future anticipation of an event important to the organism, predicted based on a previous occurrence of the same event in a similar situation (Figure 7A). In this context, the CA1 PCs receiving entorhinal cortex L3 (EC3) apical input (green line in Figure 7C) that generates plateau potentials could backproject via EC5 (blue line in Figure 7C) to attention/behaviors pertinent for responding to the event, which are memorized in CA1 and anticipatorily reactivated utilizing the predictive nature of BTSP. This could provide a mechanism for situation-dependent association of attention and behaviors voluntarily activated through EC-hippocampal loops, for instance, when the organism "intentionally" tries to associate certain behaviors with a situation. From the perspective of hierarchical information flow, this would allow the feedforward-positioned hippocampus to activate cortices lower in the hierarchy, thus closing a "loop" in processing.

Interestingly, the L1 region where MTL projections typically terminate is in lower L1 (Figure 8, panel 4), making it better positioned for easier excitation of the Ca$^{2+}$ initiation zone, compared to other CC and thalamocortical (TC) projections that terminate higher up (Figure 8, panels 1-3 and 5-7). Taking into account the MTL role in episodic memory (Rolls, 2010), it appears plausible that MTL inputs directly drive PT PCs, without the hypothesized combinatorial mechanisms used in the voluntary activation.

To focus attention on noxious and other stimuli that should not be easily habituated, such as itch, we propose that more direct links, bypassing the MTL, project from the corresponding sensory neurons to L1 tufts or other areas proximal to the Ca$^{2+}$ initiation zone of the relevant PCs.

## 5. Hypotheses on Depolarization Cross-Induction in Apical Dendritic Bundles and Minicolumn Function

Given the perspective of the proposed model on PC function, we can extend these ideas into hypotheses about minicolumn function, although this is speculative and somewhat separate from the rest of the discussion.

Neocortical PCs of layers 2/3 through 5 form minicolumns, with their apical dendrites bundling as they ascend to L1 (Figure 9A; Mountcastle, 1957; Peters and Sethares, 1996; Mountcastle, 1997). These minicolumns and bundles can be identified in all mammalian neocortical areas and are believed to serve fundamental roles in brain function, although no specific role has been definitively established. During early postnatal development, neurons within minicolumns are coupled by gap junctions, leading to coordinated neuronal calcium fluctuation patterns. However, in adulthood, these gap junctions have not been observed, and it remains unclear whether dendrodendritic synapses occur within specific bundles (Molnár and Rockland, 2020). In the bundles, the PC apical dendrites are tightly packed, with some of their membranes abutting (Peters and Sethares, 1996).

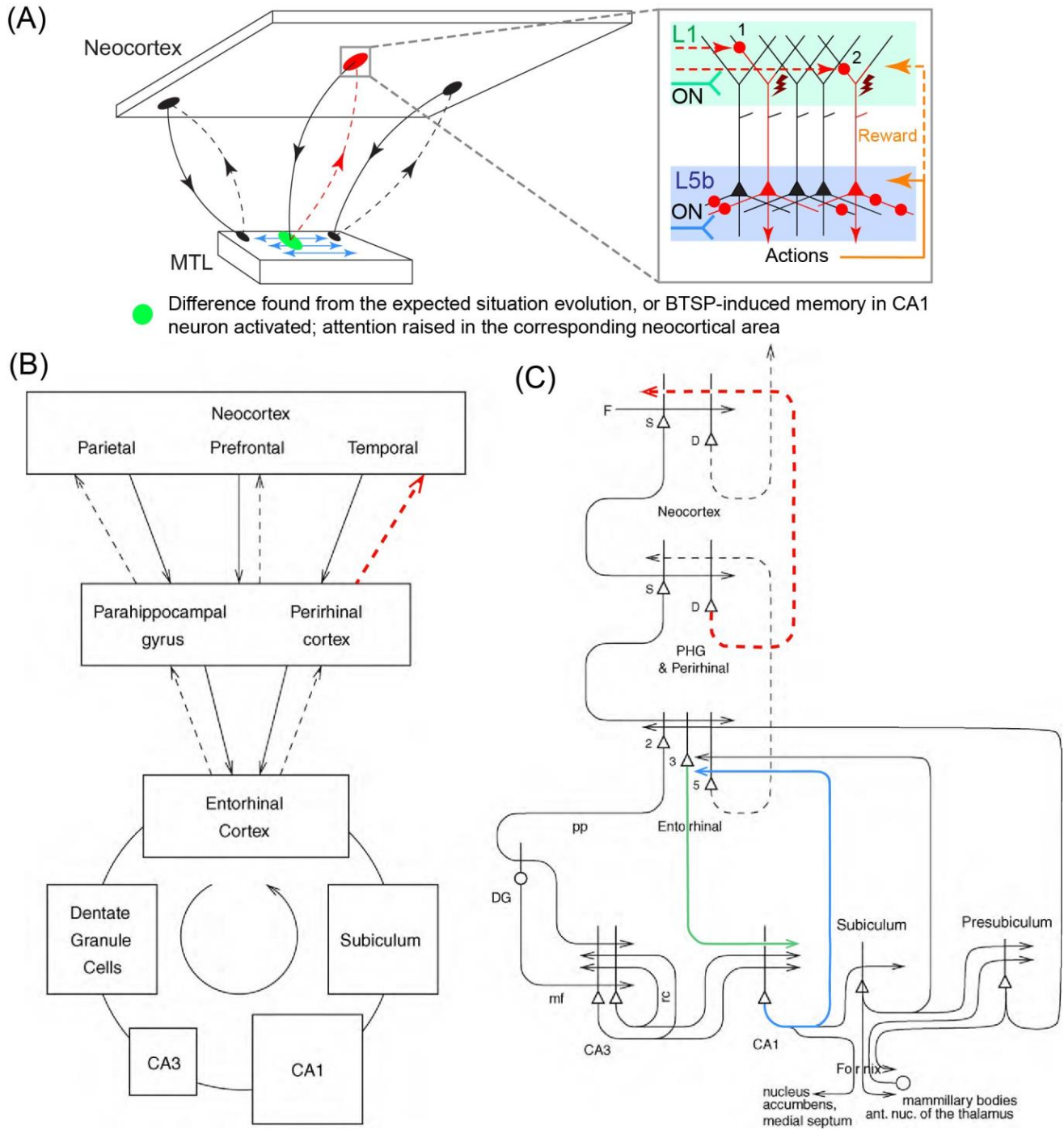

**Figure 7. Suggested mechanism for raising attention in the neocortex by the MTL. (A)** Schematic representation of the neocortical-MTL loops, with forward projections (solid lines) to the MTL and backprojections (dashed lines). Red dashed lines indicate backprojections that raise attention (red oval) in a neocortical area as a result of the detection of novelty, discrepancy, or error in the expected situation/reward (green oval), or as a result of the activation of BTSP-induced CA1 memory that predicts an important event (the same green oval). Blue arrows denote the MTL's reconstruction of episodic memories and predictions. Inset panel. Attention is raised through the sequential firing of PCs (first, labeled "1," then "2," etc.), and potentially their combinations, via L1 inputs. Other labels are defined in Figure 2. **(B), (C)** Block diagram **(B)** and a more detailed schematic **(C)** of principal pathways (solid and dashed arrows) in the neocortical-MTL loops. Backprojections (dashed lines) are thought to implement recall in neocortical areas (Rolls, 2010). The thick lines above the cell bodies represent the dendrites. Red dashed projections roughly correspond to those in **(A)**. Green and blue projections are discussed in the text. D, deep pyramidal cells; DG, dentate granule cells; F, forward inputs to areas of the association cortex from preceding cortical areas in the hierarchy; mf, mossy fibers; PHG, parahippocampal gyrus; pp, perforant path; rc, recurrent collateral of the CA3 hippocampal pyramidal cells; S, superficial pyramidal cells; 2, 3, 5 – pyramidal cells in layers 2, 3, and 5 of the entorhinal cortex, respectively. **(B)** and **(C)** modified, with permission, from Rolls (2010).

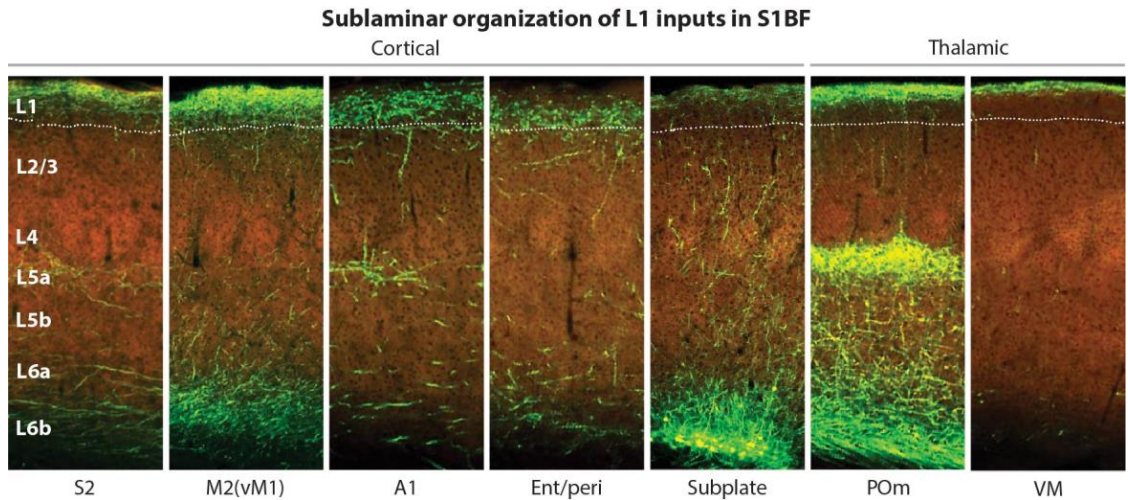

**Sublaminar organization of L1 inputs in S1BF**

Cortical — Thalamic

S2  M2(vM1)  A1  Ent/peri  Subplate  POm  VM

**Figure 8. Laminar and sublaminar distributions of corticocortical and thalamocortical L1 afferents in the mouse S1BF.** Ent/peri projections target mid/lower L1, while higher-order Pom and VM inputs are biased toward upper L1. A1, primary auditory cortex; Ent/peri, entorhinal/perirhinal cortex; M2, secondary motor cortex; POm, posteromedial thalamic nucleus; S1BF, primary somatosensory barrel field cortex; S2, secondary somatosensory cortex; VM, ventromedial thalamic nucleus; vM1, vibrissal motor cortex. Adapted, with permission, from Schuman et al. (2021). Images of fluorescently labeled projections are from the Allen Institute (http://connectivity.brain-map.org).

We hypothesize that the powerful $Ca^{2+}$ plateau potentials in some apical dendrites within a bundle can induce depolarization, via extracellular ionic flows, in the $Ca^{2+}$ initiation zones of adjacent dendrites, sufficient to affect their cell signal integration via the gain mechanism (Figure 1C), and in some instances enough to cause the dendrites to ignite plateau potentials of their own (possibly via the BAC mechanism), potentially further spreading induced depolarization throughout the bundle in a chain-like manner (Figures 9A,B). IT neurons generally have slender, "thin-tufted" apical dendrites exhibiting relatively weak branching beyond the primary bifurcation of the apical dendrite, while PT neurons are mostly "thick-tufted" with extensive branching in L1 (Harris and Shepherd, 2015). Consistent with this, we propose that L1 input primarily drives PT neurons, while IT neurons in the same minicolumn bundle "piggyback" off PT plateau potentials to drive their own learning (Figures 9 and 10).

For hierarchical L2/3 to L5b connectivity, this cross-induction mechanism could allow IT and PT neurons to collaborate, effectively creating a more hierarchical and deeply nested neuron structure (Figure 10A, panel 3). This could be aimed at circumventing the biophysical limitations on the number of basal synapses and clusters in a PC, or at prewiring certain computations.

For IT neurons projecting to other cortical areas, the cross-induction mechanism could facilitate the creation of an internal model representation that mirrors the external world (Figure 10A, panel 4). Furthermore, this mechanism could enable the prediction of outcomes of specific actions in the external world without actually executing them (Figure 10B).

Both types of connections—those that establish a hierarchical organization between different layers of the cortex and those that enable lateral communication into other cortical areas—have been observed (Douglas and Martin, 1991; Douglas and

Martin, 2004; Yamawaki et al., 2021). Interestingly, the aforementioned study on perceptual detection (Takahashi et al., 2020) reported: "The onset timings of the dendritic responses to the perceived stimuli in PT neurons preceded those in IT neurons, indicating a sequential dendritic activation across pyramidal subclasses during perceptual processing." This latency indirectly supports the hypothesis that IT neuron activations "piggyback" off PT plateau potentials.

Figures 10C,D display the results of simulations that conceptually demonstrate the benefits of prewiring some computation in L2/3, as part of the hierarchical L2/3 to L5 connectivity (Figure 10A, panel 3), and the efficiency of generating an internal representation in L2/3 of actions taken in L5 (Figure 10A, panel 4), within the dendritic cross-induction framework. Given the limited experimental data on the distribution of basal dendritic inputs, these inputs are simulated as randomly distributed. In the real world, incoming signal patterns may also have certain distributions, potentially correlated to how the corresponding synapses cluster. Since this information is not known, we model incoming signals as randomly distributed "objects" of two types: 5 adjacent active inputs ("XXXXX" configuration) and 2 pairs of 2 adjacent active inputs separated by an inactive input ("XXOXX" configuration). Furthermore, the real-world learning parameters for basal synaptic clusters are not well established; we model non-overlapping clusters such that all their inputs need to be excited for learning to occur. With these simplifying assumptions, the simulations demonstrate that the classification of patterns that differ only in the type of objects used ("XXXXX" vs "XXOXX") is substantially improved if using hierarchically connected L2/3 PCs prewired to learn to be detectors for these objects (Figure 10C). In the lateral connectivity configuration, L2/3 PCs are able to efficiently generate internal copies of the L5 learned responses (Figure 10D). One apparent limitation of this framework is that for

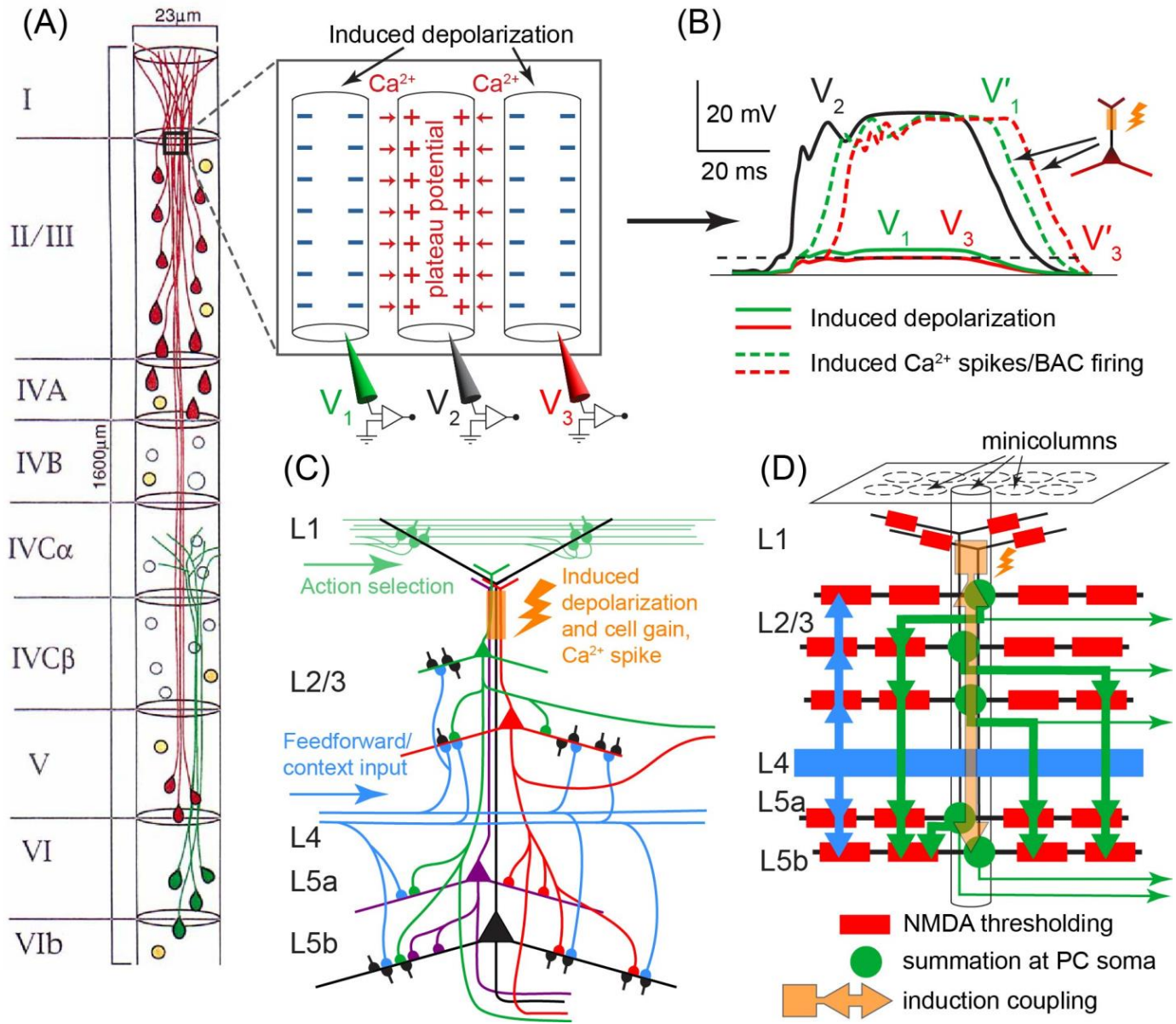

**Figure 9. Suggested dendritic plateau potential cross-induction mechanism in neocortical minicolumns. (A)** Diagram of the arrangement of neuronal somata and PC dendrites in a minicolumn (for clarity, only 50% of the cells are depicted). PCs are shown in red and green, while GABAergic neurons are shown in orange. Modified, with permission, from Peters and Sethares (1996). Inset panel. Schematic diagram of three adjacent apical dendrites in lower L1/upper L2, with the middle dendrite excited by a plateau potential that induces depolarization, via extracellular ionic flows, in the other dendrites. Red arrows depict the influx of Ca$^{2+}$ ions. The depolarization induction should be facilitated by the extended adjacency of dendrites. **(B)** Hypothetical time course of potentials in the three dendrites in **(A)**. Without tuft-soma excitatory coupling, the plateau potential in the middle dendrite ($V_2$, solid black) would induce a small depolarization in dendrites 1 and 3, shown as 15% of $V_2$ in dendrite 1 ($V_1$, solid green), and 10% of $V_2$ in dendrite 3 ($V_3$, solid red). With tuft-soma excitatory coupling and given sufficient somatic depolarization for dendrites 1 and 3, plateau potentials $V_1'$ and $V_3'$ are triggered as $V_1$ and $V_3$, respectively, cross the BAC-reduced Ca$^{2+}$ spike initiation threshold, which for simplicity is depicted at the same level for both dendrites (dashed black). **(C)** Schematic diagram of potential excitatory PC connectivity within a minicolumn and the hypothesized excitation coupling at the apical dendritic bundle (orange rectangle and flash). Only several of the possible connections are depicted. For simplicity, stellate neurons in L4 are not shown. **(D)** Schematic diagram depicting a potential hierarchy of information processing flow (thick arrows) within a minicolumn. Each PC (green round soma) symbolizes many PCs at the same hierarchical level. Dendritic branches are depicted as black lines. Layer 2 through 5 PCs learn simultaneously as they are coupled within the dendritic bundle by cross-induction from plateau potentials (orange square, arrows, and flash). Each thick green arrow signifies distributed synapsing on the basal dendritic tree of the corresponding PCs. Thick blue arrows indicate afferent basal input (feedforward and possibly some feedback context). Thin green arrows represent efferent output.

these classifications to work well with synaptic cluster sizes of 4 or more, the density of input signals has to be relatively high (>~20%), so that the clusters have a sufficiently high probability of being excited. It should be kept in mind, however, that BTSP mechanisms have been experimentally shown to rapidly and effectively express arbitrary place fields in every CA1 PC (O'Keefe, 1976; Bittner et al., 2017; Magee and Grienberger, 2020). This suggests that the basal dendritic learning machinery, which has evolved over hundreds of millions of years, and the incoming input signals are intricate to a degree that cannot be currently simulated accurately given the present state of knowledge of these processes.

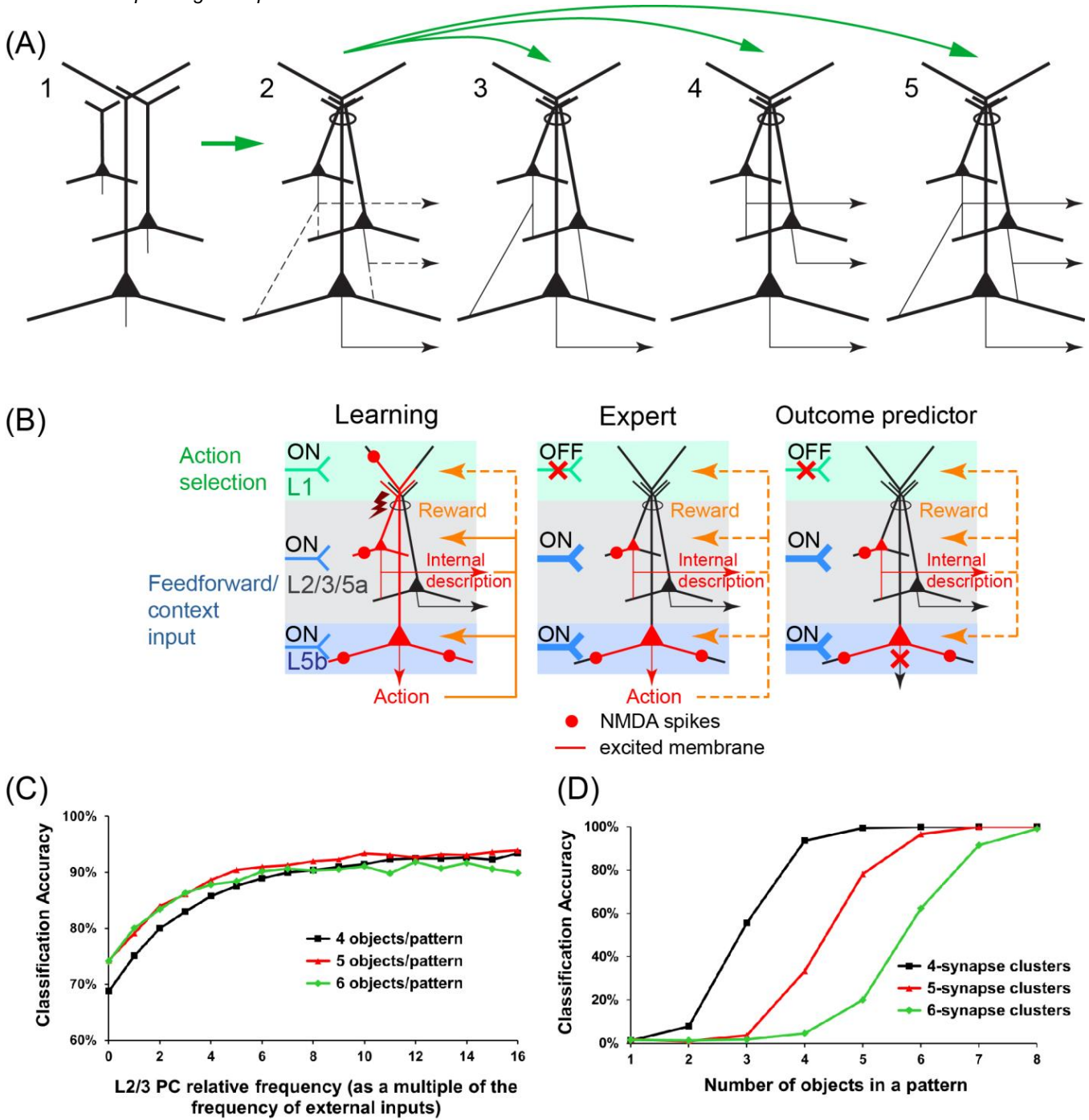

**Figure 10. Minicolumn PC connectivity enabled by the putative cross-induction of potentials in dendritic bundles. (A)** Vertically co-located PCs (panel 1) are coupled via excitation cross-induction in the apical dendritic bundles (oval in panel 2), which leaves many possibilities for basal connectivity open (dashed lines in panel 2, not all connectivity possibilities are shown). PCs in panel 2 receive similar rewards and learn simultaneously, which can slow the learning process but also increase complexity. Panels 3-5. Various realizations of the diagram in panel 2. Panel 3. The depicted connectivity can functionally establish a pyramidal "hyperneuron" capable of circumventing the physical limitations on the number of synapses and clusters in an actual PC. PCs in the upper levels can learn or be prewired to encode frequently occurring and/or important input combinations, which are further processed as inputs in the lower levels. Panel 4. PCs in some layers can learn to mimic actions taken by PCs in other layers for specific input combinations, to possibly further process/predict the actions' effects in an internal representation forwarded downstream. Panel 5. The combination of connections in panels 3 and 4 (the same as the connections depicted in panel 2) can integrate their functionalities. **(B)** Elaboration on the potential learning process in the configuration in **(A)**, panel 4. Learning (left panel) and having learned (middle panel) a beneficial motor action or classification. Right panel. Internal copies of the intended action can be forwarded for internal predictive processing without actual execution of the action if the output from the lower PC is inhibited. Notation is defined in Figure 2. Design of **(B)** is inspired by Doron et al. (2020). **(C), (D)** Simulation of learning in 10 minicolumns, each with 1 L5 PC receiving 100 external inputs that randomly synapse on the PC's basal dendritic trees. There are 100 different input patterns (10 patterns per minicolumn) and 20,000 basal synapses per L5 PC. Basal synapses form random, non-overlapping clusters that increase in weight given input co-activation, the PC firing activated via its apical dendrite, and a positive reward. One learning episode per input pattern. Given an input pattern, the minicolumn with the largest L5 output is the one that fires. **(C)** The configuration in **(A)**, panel 3. Basal synaptic cluster size is 4. Two L2/3 neurons are prewired to learn to identify objects in the "XXXXX" and "XXOXX" configurations (see text for description). The L2/3 PCs synapse on their minicolumn's L5 PC basal dendritic trees with a frequency specified as a multiple of the frequency of the external inputs, shown on the x-axis (0 frequency corresponds to no L2/3 to L5 connectivity). Learning with 50 different random non-overlapping and non-touching "XXXXX" input patterns and 50 corresponding similarly positioned "XXOXX" patterns. The accuracy of classification improves with increasing L2/3 to L5 connectivity. **(D)** The configuration in **(A)**, panel 4. One L2/3 PC per minicolumn, similar to the L5 PC (20,000 basal synapses and the same cluster size), is not connected to it basally and learns in parallel. Input patterns are objects in the "XXXXX" configuration. The accuracy of simultaneous correct classification in L2/3 and L5 is above 98% for a sufficiently high number of objects (e.g., 5 or more objects for synaptic cluster size 4). The details of the simulations are presented in the Materials and Methods section.

In the present framework, the primary function of the cortex is to classify its inputs, which is ultimately possible due to the inherent classifiability of our world. This classification task is divided into smaller (and likely more biophysically tractable) subtasks by the various cortical areas and their interconnections, in a way that attempts to mirror the classifiability structure of the external world in its relation to the internal world. Within individual cortical areas, minicolumns function as basic classifier units, responding to combinations of basal inputs spanning hundreds of micrometers according to hardwiring and memory from learning. Within minicolumns, co-excitation via dendritic bundles allows pyramidal neurons in different layers to collaborate, to create a "synthetic hyperneuron" that is more powerful and hierarchical, and to generate an internal representation of the external world, which can be used to predict the outcomes of an organism's actions. It seems likely that, given the putative dendritic coupling, more complex functional roles of minicolumns might have been achieved.

## 6. Efficient Trial Action Selection Through Tuft-Soma Excitation Coupling

We suggest that the tuft-soma excitation coupling (Figure 1) enables the preferential selection of PCs with stronger basal input for trial firings, promoting learning in accordance with the hierarchical connectivity of the cortex and any hardwired predispositions for specific learning. Without this coupling, an organism might waste time sampling actions that are less learnable or relevant. Subjectively, this coupling may be experienced as "intuition," such as when guessing an action to solve a problem. On the other hand, higher-order thalamic projections to the cortex, such as those from the posteromedial nucleus (POm) to both L1 and L5a (Figure 8), may be able to drive PT PCs to bursting with dendritic plateau potentials (Suzuki and Larkum, 2020) regardless of their other inputs.

## 7. A Larger-Scale Picture

As a microcircuit framework, the present model can be incorporated within larger-scale frameworks such as the Global Neuronal Workspace (GNW; Baars, 1988; Dehaene and Changeux, 2011), Default Mode Network (DMN; Raichle, 2015), and Integrated Information Theory (IIT; Tononi et al., 2016). In the model, information in the neocortex dynamically evolves through PC $Na^+$ spike trains, based on prewired and learned connectivity, with PC burst firing with dendritic plateau potentials superimposed via voluntary and involuntary attentional pathways. In a wakeful and quiet state, we suggest that a minimal amount of bursting with plateau potentials is maintained to impose active situation exploration and learning (as opposed to a passive response to incoming stimuli), as the focus of attention is automatically directed toward the most salient "background" stimuli and percepts. This can be achieved by uniformly augmenting PC excitations until a sufficient number reach plateau potentials/bursting. This process could be accomplished by adjusting the apical-basal coupling strength using projections from higher-order thalamic areas such as Pom (Suzuki and Larkum, 2020), or by other processes similar to the "action initiation" mechanism described earlier. This is consistent with the idea that decoupling of apical and basal information streams leads to a loss of consciousness (Suzuki and Larkum, 2020), which we interpret as being associated with awareness and attention.

Developing a network-scale theory based on the ideas presented in this paper is beyond the scope of this work. However, potential building blocks for such a theory might include: the ability to memorize factual knowledge, such as what word corresponds to an ensemble of active PCs or minicolumns representing an object or concept (Figure 11A), and how to respond to a sequence of situations (Figure 12); conceptual knowledge, such as understanding principles, theories, and models (implemented at the cellular/minicolumn level as classifications, Figures 2D, 3, and 11B); and the ability to form an internal predictive representation of external events, potentially using IT PCs (Figure 10).

In one of these implementations, individual minicolumns or PCs may represent basic features that are co-activated to represent a concept (Figure 11B). Various flavors of a concept (such as those expressed by synonymous words) might be represented by some features that are not co-activated with the commonly co-activated features. In a sentence, the relationship between parts of speech such as nouns, adjectives, and verbs should reflect the processes they describe in the world (Figure 11C), with the typical sentence length potentially reflecting the capacity of working memory. The ability to imagine an arbitrary concept, such as a "pink flying elephant," is then possible through the co-activation of the concepts representing each of the component words. However, this concept is unlikely to be remembered or frequently used, as it lacks utility.

Solving a problem could proceed by identifying a feature that was not previously considered (or paid attention to) that leads to the achievement of a goal (Figure 12). This may include identifying a feature common to another process for which the solution is known. This success then generates a positive reward, which triggers the memorization of the solution (such as memorization of a new classification, Figure 2), and, given substantial rewards, may result in the formation of an associated hippocampal episodic memory (compare to a somewhat opposite role of the hippocampus in Aru et al. (2023)). Interestingly, the term "analysis", which literally translates from Greek as "breaking up," implies decomposing a problem into its basic features, the relationships among which can be studied. This aligns with the present framework on how concepts are stored. A practical challenge for an organism arises when real-world problems have numerous interacting elements.

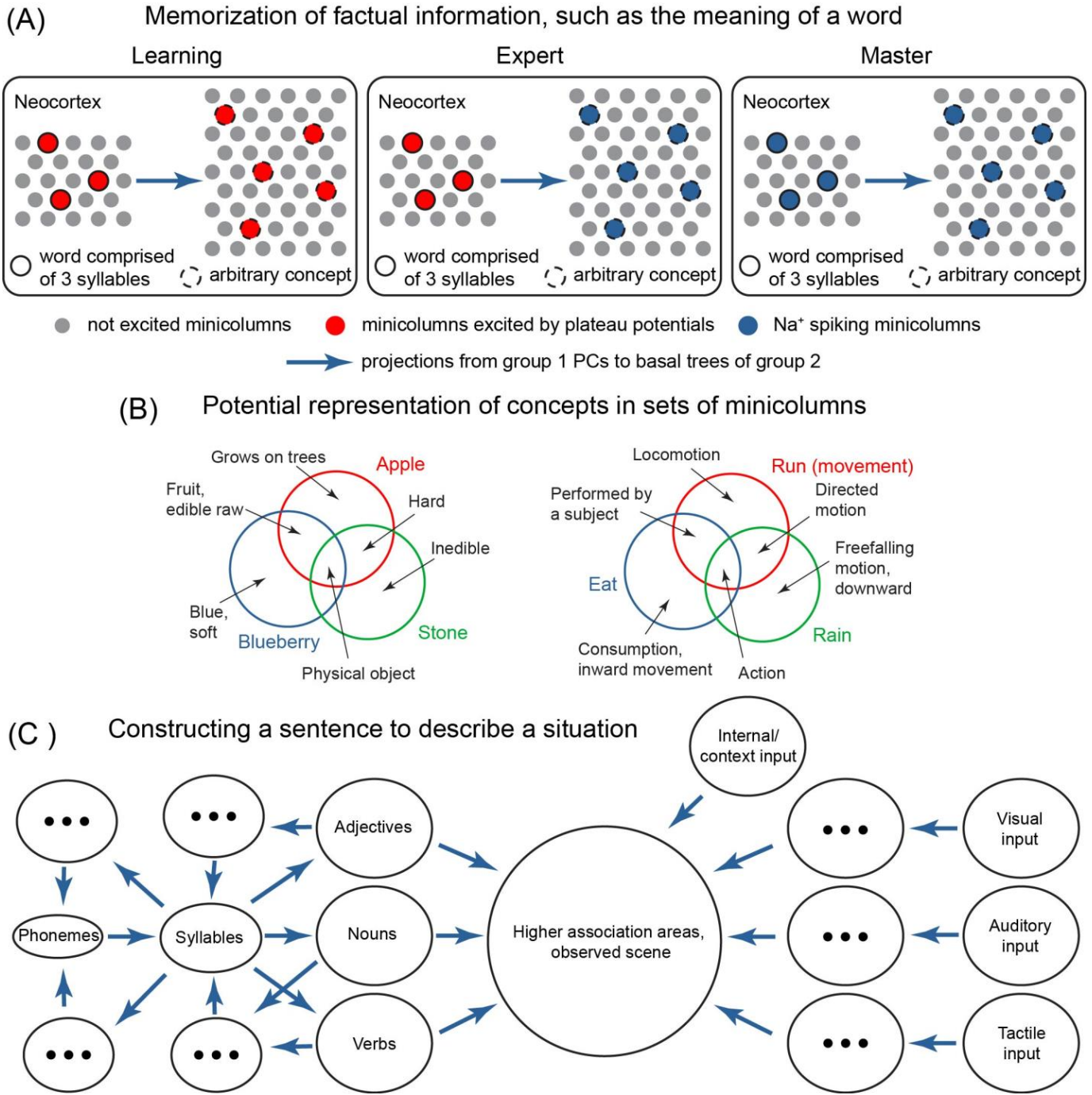

**Figure 11. Several potential building blocks for a network-level theory. (A)** Schematic representation of a neocortical connectivity configuration implementing the learning of a word's meaning. In the panels, the left group of minicolumns projects to the basal trees of the group on the right, which can be either smaller or larger. The left group represents syllables that comprise words, the right group represents arbitrary concepts. To associate a word with a concept, both are (voluntarily) activated via L1 inputs, and a positive reward (not depicted) is generated either internally or with the help of an external critic, such as a teacher, to acknowledge successful activation (left panel). After learning, attentional activation of the word elicits the concept (middle panel), using the combinatorial memory mechanisms described earlier. After sufficient practice, the word activation can become inattentive and still continue to elicit the concept (right panel). For example, the 3-syllable word could be "blueberry," and the 5 active minicolumns in the right group could represent the concepts "round," "blue," "small," "sweet," and "slightly sour," which are activated when a blueberry is visualized. Episodic memories related to blueberries may be concurrently activated in the hippocampus. Basal synaptic clustering may involve certain optimizations, such as excluding combinations that do not co-occur, such as "blue" and "white." Instead of words and their meanings, the left group could represent letters, and the right group could represent key pressing in learning how to touch type, or other association learning tasks. Alternatively, the minicolumns in the diagram could represent individual PCs. **(B)** Schematic representation of how concepts corresponding to various words could be represented in ensembles of minicolumns. The circles represent the ensembles of minicolumns activated by the concepts (text in color), while more basic concepts (black text and arrows) are represented in the minicolumns shared between the bigger concepts, as indicated. This type of representation provides inbuilt storage of classifications. The bigger concepts can be updated by including or excluding component minicolumns, which could be located in various cortical regions. **(C)** Suggested process for formulating a verbal description of a perceived scene. Sensory feedforward pathways (ovals on the right) combine with internal and context input (top right oval) in higher association areas (large circle). Trial sentences are mentally composed (ovals on the left) and compared for their ability to describe the scene (large circle), until a satisfactory sentence is constructed. The language rules for combining parts of speech mirror the general structure of the processes in the world, similar to the description in Langacker (1987). Blue arrows denote projections from one group of minicolumns to the basal trees of another. Concepts evoked by combinations of minicolumns can be used to reconstruct the minicolumns that evoked them via reverse paths involving more cortical areas (backprojecting blue arrows).

## Learning arbitrary responses to arbitrary sequence of situations

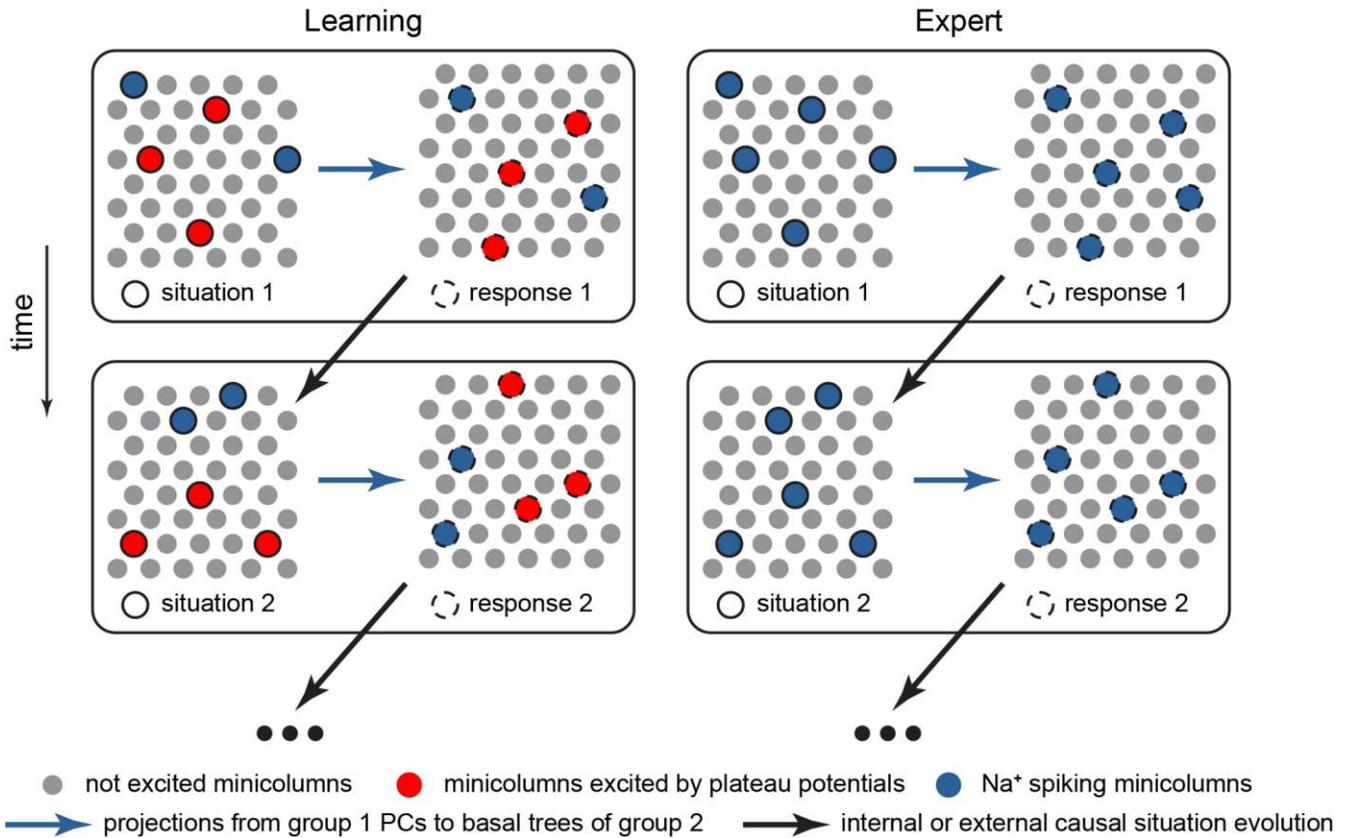

**Figure 12. Schematic representation of a potential neocortical connectivity that implements trial-and-error learning of responses to dynamically evolving situations.** The left group represents features of observed situations, the right group represents possible responses. Within the left group, the minicolumns bursting with dendritic Ca$^{2+}$ spikes (in red) are the situation features that are voluntarily and arbitrarily attended to, leading to their more pronounced output compared to the Na$^+$ spiking minicolumns (in blue). In the right group, Na$^+$ spiking minicolumns (in blue) are automatically executed and do not learn, while the set of minicolumns bursting with dendritic Ca$^{2+}$ spikes (in red) is voluntarily and intentionally manipulated, potentially by the cortices deep in the hierarchy, until a positive reward (not depicted) is achieved, possibly with assistance from an external critic. Both groups may be distributed across various cortical areas, with hierarchical learning (not depicted). Paying attention to a previously disregarded aspect of a situation for which the optimal response is already known, which can involve different cortical areas and levels of abstraction (e.g., Isaac Newton focusing on the relative motion between a falling apple and the Earth, and his conception of forces acting between two objects), may lead to the achievement of a goal and the generation of a reward, resulting in the memorization of the solution. The modular structure, in which situation features allow for classification, stems from the inherent classifiability of our world. The situation may evolve in time (the vertical change in the situation and responses). Alternatively, the minicolumns in the diagram could represent individual PCs.

## 8. Discussion

We have built upon the neocortical PC learning paradigm from Rvachev (2013), in which apical tuft or perisomatic inputs initiate "guess" PC firings, while excited basal synaptic clusters encode input patterns, with the cluster excitation strength adjusted based on reward feedback. We have modified the cluster learning rules to be quite similar to the experimentally observed BTSP (Bittner et al., 2017), while also assuming that negative rewards (punishments) reduce cluster excitation strength, as in Rvachev (2013). Even without this assumption, the framework could perform well by solely strengthening and homeostatically renormalizing synaptic weights. However, memory weakening via punishment is an appealing mechanism for the synthesis of abstract concepts (Figure 3), and it is a model prediction. Another departure from BTSP is that place cells, which we have hypothesized drive an attention signal, entirely retune their response to a new situation after re-potentiation by a dendritic Ca$^{2+}$ spike (Diamantaki et al., 2018; Milstein et al., 2021). In contrast, our framework suggests that neocortical PCs accumulate learned weights, although one-shot learning is also possible. We hypothesize that the retuning of place cells might be behaviorally necessary, given their proposed role in attention. In neocortical PCs, however, we believe multiple input patterns should be classifiable into the same output. While the time-asymmetric predictiveness of BTSP is not necessary in our model, it is not excluded either. The differences from BTSP are quantitative rather than qualitative and are predictions of our model. Therefore, the demonstrated effectiveness of BTSP in rapidly expressing arbitrary place fields in every CA1 PC (O'Keefe, 1976; Bittner et al., 2017; Magee and Grienberger, 2020) supports the presented "combinatorial switching" framework. On the other hand, the numerical results from Rvachev (2013) remain valid in the updated formulation, particularly the surprisingly intelligent classification behavior and strong performance in memorization tasks.

Several features of the model, namely, reinforcement learning driven by the combination of pre-synaptic activity, post-synaptic apical tuft $Ca^{2+}$ entry, and neuromodulators, as well as the link of the $Ca^{2+}$ entry to attention, are similar to those in Roelfsema and Holtmaat (2018). However, in their model, attention is shifted to PCs that caused an action, via feedback connections from the PCs performing the action, while in our model, the $Ca^{2+}$ spiking neurons are performing the action in the first place and are therefore the initial subject of attention. As was shown in Rvachev (2013), the system does not require feedback connections to PCs that caused an action to function efficiently and learn useful abstract concepts. Therefore, our model, in a certain sense, represents a simplification of the mechanisms described in Roelfsema and Holtmaat (2018).

Our model differs from other current microcircuit-level theories of neocortical function. In Larkum (2013), the cortical association between the external and internal worlds is accomplished through the integration of PC apical and basal streams. While the anatomy of feedforward/feedback connections and the apical-tuft PC excitation coupling have been experimentally established (Felleman and Van Essen, 1991; Larkum et al., 1999), the hypothesis inherent in Larkum (2013) that L1 receives the internal world representation has not yet been experimentally confirmed. In our model, the association between the external and internal worlds is achieved implicitly through $Na^+$ spike train processing (possibly both bursting and regular) of the external stream via modified basal connections, coupled with the interaction with the MTL and subcortical structures. Therefore, the perception of the external world is automatically shaped by prewired and learned connections.

As discussed in Section 4.3, our model also substantially differs from the predictive coding theories (Keller and Mrsic-Flogel, 2018), in that the error correction signal is global, similar to Roelfsema and Holtmaat (2018), rather than local. While many signals consistent with local prediction errors have been observed, they are also consistent with other phenomena, and cortical neurons explicitly coding for prediction error have not been experimentally identified (Keller and Mrsic-Flogel, 2018). On the other hand, the global reward signals of the kind used in our model are well established (e.g., Schultz (2017)).

Several other features of our model also differ from existing formulations and can explain a range of observed behaviors. The PT PC $Na^+$ spike train processing is automatic (inattentive) and is interspersed with dendritic $Ca^{2+}$ spikes with burst firing, which are subjectively perceived as attentional/aware processing (and are likely accompanied by inbuilt attention-focusing mechanisms). These dendritic $Ca^{2+}$ spikes are initiated via apical tuft inputs, either voluntarily and arbitrarily (using "free will"), presumably through NMDA spikes elicited via CC and CTC afferents, or involuntarily, via MTL and subcortical afferents directing attention towards important cues such as novelty, discrepancy, and noxious or unexpected stimuli. This formulation aligns with the observation that learned behaviors/classifications in mammals become automated, while attention is generally required for learning and can be directed either voluntarily or involuntarily. In line with this, we have classified PT PC excitations as attentional versus automatic and voluntary/acquired versus involuntary, based on their excitation pathway. Voluntary attentional actions become basally-executed acquired automatic actions with learning, while involuntary attentional actions do not become automatic due to their hardwiring to the apical inputs.

We have proposed that the dual arborization of the PC dendritic tree elegantly permits "free will" to coexist with the internal representation of the external world, by allowing the initiation and control of arbitrary actions from various (including deep) cortical areas via apical tufts, while simultaneously offering access to, and the potential for refinement of, the external world representation stored in the basal trees. This basal tree representation, distributed over cortical areas, is reminiscent of the structure of artificial neural networks. However, these networks do not learn classifications in the intermediate layers via trial-and-error dendritic learning. Therefore, our model provides guidance for a new architecture. In this context, it should be noted that in a biological setting, "supervised" learning with a teacher is essentially trial-and-error, as the organism follows learning examples.

Within our framework, the primary function of the cortex is to classify its inputs to suit the organism's needs, a task that is, in principle, achievable due to the inherent classifiability of our world. Motor actions and behaviors form some of these classifications. This large classification task is divided into smaller subtasks by the various cortical areas and their interconnections, in a way that attempts to mirror the classifiability structure of the external world in its relation to the internal world. Within individual cortical areas, we have hypothesized that minicolumns function as basic classifier units, responding to combinations of basal inputs according to hardwiring and memory from learning. Within these minicolumns, we have proposed that co-excitation via depolarization cross-induction within dendritic bundles allows pyramidal neurons in different layers to collaborate, to create a "pyramidal hyperneuron" that is more hierarchical and powerful than an actual PC, and to generate an internal representation of the external world, which can be used to predict the outcomes of an organism's actions. We have demonstrated via simulations that these mechanisms are conceptually possible. This putative co-excitation mechanism may have enabled more complex functional roles for minicolumns, possibly including temporal signal processing. Furthermore, we have hypothesized that intelligence similar to ours (i.e., classification-based) might exist in other universes (Tegmark, 2015) with a large number of elementary particles interacting under laws invariant in space and time (or other analogous dimensions).

Besides the biophysical plausibility of the model's plasticity mechanisms noted above, the presented framework agrees with experiments that showed that learning to associate electrical microstimulation of S1 L5 PCs with a reward, but not for sensory perception (Doron et al., 2020). In our interpretation of the experiment, for sensory stimulation to attract attention and become available for learning, the MTL must recognize it as a novel stimulus and direct attention to it via perirhinal input. Further, there is some evidence suggesting that the feedback input to L1 can be driving (Mignard and Malpeli, 1991; Covic and Sherman, 2011; De Pasquale and Sherman, 2011). Regarding the hypothesis on minicolumn depolarization cross-induction, the observed latency of IT PC dendritic activations relative to PT PC (Takahashi et al., 2020) indirectly supports the idea that IT activations are dependent on PT. More direct experimental verification is desirable.

As a microcircuit model, the presented model can be incorporated within larger-scale frameworks such as GNW, DMN, and IIT (Dehaene and Changeux, 2011; Raichle, 2015; Tononi et al., 2016), and it conceptually aligns with the dendritic integration theory (Bachmann et al., 2020) and a generic optimization approach such as the minimization of free energy (Friston, 2010). The expansion of this model to encompass network scales is beyond the scope of this work. Nonetheless, we have outlined several potential building blocks: the network for memorizing factual knowledge, such as the meaning of a word (Figure 11A), and dynamic response to evolving situations (Figure 12); the architecture for acquiring and storing conceptual knowledge, such as the representation of concepts through classifications in a single PC as well as in ensembles of PCs and minicolumns (Figures 2D, 3, and 11B); and the capability, in principle, to create an internal predictive representation of external events (Figure 10). Language appears to be particularly suitable for demonstrating cortical classification operations, and we have presented a potential implementation of word semantic storage and sentence construction (Figures 11B,C). We have also suggested that dynamic focusing of attention on different aspects of input, together with trial-and-error, is used to solve a problem (Figure 12). These network implementations are basic examples, and given the proposed microcircuit tools, evolution likely has conceived more sophisticated solutions. As a general paradigm, however, it seems plausible that higher-order cortical areas dynamically govern overall behavior via L1 projections while receiving feedback from the external world and internal sources (e.g., MTL), and depending on the situation, utilize one of the mechanisms listed above, as well as others.

The work we have presented outlines general principles and mechanisms that allow for a wide range of implementations, many details of which remain uncertain. For example, it is not clear whether voluntary behavior is driven by substantially all CC and CTC L1 projections or by a small subset of one or both; how IT PC activity relates to attention, awareness and voluntary actions; which mode of PT PC firing drives attentional versus automatic behaviors (e.g., as suggested, dendritic $Ca^{2+}$ spiking with burst firing, or perhaps also less intense $Na^+$ burst firing without dendritic $Ca^{2+}$ spikes, as in the recall-associated bursts in hippocampal place cells); and whether some of the excitations hypothesized to proceed via L1 input can also proceed via other apical inputs proximal to the $Ca^{2+}$ initiation zone. It should be noted that while it is quite likely that at least some of the ideas presented here will turn out to be incorrect, many hypotheses are independent.

Further questions include: How are ensembles of excited PCs stabilized and synchronized within various brain rhythms? Does PT PC output from different cortical areas trigger fundamentally different behaviors, or a standardized set corresponding to behavioral classifications such as "feeding opportunity," "threatening," "approved by teacher/society," etc.? What dynamic interactions, other than those already discussed, emerge within and between cortical and subcortical areas, and how do they relate to behaviors such as recognizing something familiar, or finding something interesting or funny? What is the functional basal cluster size in the PC, and how does it vary across cortical areas? Does the number of individual phonemes and syllables in a language reflect the combinatorial processing capabilities of minicolumns and PCs? Can a hierarchical storage structure similar to that in Figure 11B be used throughout the entire cortex? How is BTSP affected by neuromodulators? How does homeostatic renormalization of synaptic strength function? Does intracellular pressure integration enhance PC clustered input detection (Rvachev, 2003; 2010; 2013)? Is classification within various cortical areas related to learning using statistical invariants (Vapnik and Izmailov, 2020)? How does temporal processing proceed (see Rvachev (2012) for some interesting ideas)?

In summary, we have proposed novel operating principles and cellular mechanisms that plausibly describe cortical function. Although many details are yet to be uncovered, we hope this can serve as a guiding framework for future experimental and theoretical research in this direction.

## 9.    Materials and Methods

The simulations for Section 5 were performed using a Perl program, modified from Rvachev (2017). The simulations consisted of several stages described below.

*Overall configuration.* All inputs and outputs operated in an "on" or "off" regime. Each of the 10 simulated minicolumns contained one L5 PC. In the "hierarchical connectivity" configuration (Figures 10A, panel 3 and 10C), two L2/3 PCs per minicolumn received external inputs and fed their output into their minicolumn's L5 PC basal dendritic tree. In the "internal representation" configuration (Figures 10A, panel 4 and 10D), one L2/3 PC per minicolumn was simulated.

*Generation of synaptic clusters on L5 PC basal dendritic trees.* For each run and each L5 PC, non-overlapping synaptic clusters of size 4, 5, or 6 were randomly generated from 100 external inputs and two L2/3 inputs for the hierarchical configuration. The simulation used uniform distribution sampling, with the sampling space size the same for each of the external inputs, but scaled up for the two L2/3 inputs by a factor F, shown on the x-axis in Figure 10C. Clusters were not allowed to have two or more synapses from the same input. Identical clusters were allowed. The limit was 20,000 basal synapses per PC.

*L2/3 PCs.* For the hierarchical configuration, one of the L2/3 PCs learned to respond with output only when an "XXXXX" input was present (any 5 adjacent inputs active), the PC's apical dendrite was excited through the dendritic bundle depolarization cross-induction, and a positive reward was observed. The other L2/3 PC learned similarly, but only when an "XXOXX" input was present (any 2 pairs of adjacent inputs active, separated by an inactive input). The basal dendritic trees for these PCs were not explicitly simulated; however, these responses could be trivially reproduced with trees having 96 clusters of 5 (excitatory and inhibitory) synapses, for the 100 inputs. For the internal representation configuration, the simulated L2/3 PCs were similar to the L5 PCs (20,000 synapses per basal dendritic tree and the same cluster size).

*Generation of input patterns.* Input patterns consisted of all objects of either "XXXXX" or "XXOXX" type. The objects were not allowed to overlap or touch (minimum spacing between objects of 1 input). Each minicolumn was assigned 10 of 100 generated patterns. For the hierarchical configuration, for each of 50 different generated random "XXXXX" patterns, a corresponding "XXOXX" pattern with the same object locations was assigned to a different minicolumn. If this correspondence of object location between patterns was not enforced, the overall classification accuracy was higher and rather similar for various values of F. For the internal representation configuration, the randomly generated patterns were of the "XXXXX" type.

*Training.* Each pattern was presented to the system once, with the assigned L5 PC fired via its apical input (with the dendritic bundle cross-induction exciting the adjacent L2/3 apical dendrites), and a positive reward delivered to the system. A cluster's weight was incremented by 1 if all its synapses were active, its PC fired, and a positive reward was observed.

*Testing.* All training patterns were presented for testing, one at a time. For each pattern, the weights of clusters that had all their synapses active were summed for each L5 PC. The L5 PC with the largest sum of weights was the only one that fired.

If two or more L5 PCs had the same weight, one of them was selected randomly.

*Averaging.* For the hierarchical configuration, results from 50 independent runs were averaged. For the internal representation configuration, results from 10 independent runs were averaged.

## 10.    Conflict of Interest

The authors declare that the research was conducted in the absence of any commercial or financial relationships that could be construed as a potential conflict of interest.

## 11.    Author Contributions

MR devised the theory and wrote the manuscript.

## 12.    Acknowledgments

The author thanks Michael A. Rvachov and Timur M. Rvachov for many invaluable insights and discussions.